%% file: sample-sigconf-authordraft.tex
\documentclass[acmsmall]{acmart}

\AtBeginDocument{%
  }

\setcopyright{cc}
\setcctype{by}
\acmJournal{PACMHCI}
\acmYear{2026} \acmVolume{10} \acmNumber{6} \acmArticle{CSCW149}
\acmMonth{10} \acmDOI{10.1145/3816997}

\usepackage{booktabs}
\usepackage{caption}
\usepackage{subcaption}
\usepackage{graphicx}
\usepackage{array}
\usepackage[export]{adjustbox}

\begin{document}

\title[Shaping Collaborations with Algorithms]{Shaping Collaborations with Algorithms: How Agency and Heterogeneity Criteria Influence Team Formation and Outcomes}


\author{Diego Gomez-Zara}
\email{dgomezara@nd.edu}
\orcid{0000-0002-4609-6293}
\affiliation{%
  \institution{University of Notre Dame}
  \city{Notre Dame}
  \state{IN}
  \country{USA}
}

\author{Victoria Kam}
\affiliation{%
  \institution{Northwestern University}
  \city{Evanston}
  \state{IL}
  \country{USA}
}

\author{Charles Chiang}
\affiliation{%
  \institution{University of Notre Dame}
  \city{Notre Dame}
  \state{IN}
  \country{USA}
}

\author{Jiarui Xia}
\affiliation{%
  \institution{University of Notre Dame}
  \city{Notre Dame}
  \state{IN}
  \country{USA}
}

\author{Ploenta Voraprukpisut}
\affiliation{%
  \institution{Northwestern University}
  \city{Evanston}
  \state{IL}
  \country{USA}
}

\author{Leslie DeChurch}
\affiliation{%
  \institution{Northwestern University}
  \city{Evanston}
  \state{IL}
  \country{USA}
}

\author{Noshir Contractor}
\affiliation{%
  \institution{Northwestern University}
  \city{Evanston}
  \state{IL}
  \country{USA}
}

\renewcommand{\shortauthors}{Gomez-Zara et al.}

\begin{abstract}
\input{00_Abstract}
\end{abstract}

\ccsdesc[500]{Human-centered computing~Empirical studies in collaborative and social computing}
\ccsdesc[100]{Applied computing~Psychology}
\ccsdesc[100]{Theory of computation~Algorithmic game theory and mechanism design}
\ccsdesc[500]{Information systems~Collaborative and social computing systems and tools}
\ccsdesc[500]{Human-centered computing~Collaborative and social computing theory, concepts and paradigms}

\keywords{Team Formation, Team Heterogeneity, Augmented Intelligence, Heterogeneity Criteria, Lab Experiments, Recommendation Systems}

\received{May 13, 2025}
\received[revised]{January 13, 2026}
\received[accepted]{March 17, 2026}

\maketitle

\input{01_Introduction}
\input{02_Literature_Review}
\input{03_Methodology}
\input{04_Results}
\input{05_Discussion}
\input{06_Conclusion}

\begin{acks}
  This publication resulted from research supported by the National Science Foundation (SES-2021117, SES-2341431, SES-2341432), Microsoft Research (2020 Microsoft Research Dissertation Grant), Alfred P. Sloan Foundation (G-2024-22427), and Amazon Science (2023 Research Award). The funders had no role in study design, data collection and analysis, decision to publish, or preparation of the manuscript.
\end{acks}

\section*{Disclosure of Generative AI Use}
The authors used generative AI tools  (e.g., ChatGPT, Grammarly, and Claude) to assist with manuscript editing. These tools were not used to generate data, conduct analyses, or produce results. The authors reviewed all AI-assisted text and take full responsibility for the manuscript.
\bibliographystyle{ACM-Reference-Format}
\bibliography{sample-base}


\input{07_Appendix}

\end{document}

%% file: 00_Abstract.tex
Across professional networking platforms, scientific collaboration networks, co-founder matching tools, and workplace collaboration platforms, algorithms increasingly shape how individuals find, evaluate, and connect with potential collaborators. These systems create tensions between user agency and organizational values: Should algorithms organize individuals directly in line with organizational goals? Should algorithms allow individuals to choose freely? Or should algorithms subtly nudge choices toward those goals while preserving user agency? Each approach has implications for who gains access to collaborators, opportunities, and professional networks. This study examines how team formation algorithms that vary in user agency and incorporate organizational values---specifically, promoting teams with different expertise and backgrounds---influence collaborator selection, team composition, team processes, and team outcomes. We conducted a $2 \times 2$ between-subjects laboratory experiment using a team-formation recommendation system, manipulating user agency (assignment vs. choice) and heterogeneity criteria (included vs. not included). Across four experimental conditions, 332 participants either selected collaborators through the system or were assigned to teams by the system, and then worked as members of the resulting 83 teams. Results show that modest differences in algorithm design can systematically reshape team composition and collaboration decisions, often without users fully perceiving the system's influence. While allowing user agency reinforced homophily, nudging by reordering recommendations based on heterogeneity criteria increased the selection of different collaborators and produced teams that performed better than those formed through unconstrained choice. Nevertheless, nudging operated without users' awareness, raising questions about transparency and autonomy in collaborative systems. Our findings demonstrate that algorithms embedded in collaboration platforms constitute a distinct mode of algorithmic governance, where resolving tensions between user agency and organizational values raises fundamental questions about transparency, access, and control over collaboration. We discuss how these algorithmic designs become social gatekeepers by shaping social opportunities and networks rather than directly controlling tasks or performance.

%% file: 01_Introduction.tex
\section{Introduction}
\label{sec:introduction}
In the modern workplace, algorithms increasingly reshape how individuals find, evaluate, and connect with potential collaborators \cite{Polzer2022-pl,Kellogg2020-oo,Laapotti2022-zf,Jain2021-ef,Faraj2018-tz}. Across professional networking platforms, scientific collaboration networks, co-founder matching tools, and workplace collaboration platforms, these computational artifacts can recommend new teammates, rank individuals, or prioritize existing relationships \cite{De-Arteaga2019-ah, Gomez-Zara2020-ol, Helberger2018-tt}. While individuals use these technologies to search for collaborators, organizations can deploy algorithms that encode organizational values into individuals' searches, thereby guiding decisions about whom to work with. These algorithmic designs can also unintentionally exacerbate social biases, guide users' behavior without transparency, reduce the visibility of some potential collaborators, and narrow users' collaboration opportunities \cite{Glaser2021-cc, Nedzhvetskaya2024, De-Arteaga2019-ah}. As such, algorithms increasingly function as \textit{social gatekeepers}, shaping visibility, access to collaboration opportunities, and the pathways through which professional relationships form. Rather than merely supporting coordination, these systems actively mediate who gets to work with whom---and under what conditions---producing downstream consequences for team composition, access to collaboration, and professional opportunity \cite{Faraj2018-tz,Selbst2019,gillespie2014relevance}.

Previous research has examined the consequences and trade-offs of employing algorithms to enhance team formation \cite{Gomez-Zara-2020-cscw, Barnabo2019}. These systems embed different assumptions about user agency and organizational values. On the one hand, algorithms can assign members to work together based on organizational criteria, such as competence or team composition. Although these algorithms can promote collective goals, they may reduce individuals' engagement with their team due to perceived risks associated with working with unfamiliar people or people from other groups \cite{Kaiser2013-wa, Dobbin2016-xt, Dover2016-xx}. On the other hand, algorithms that allow individuals to choose their collaborators can increase their commitment and satisfaction with the assembled team \cite{Chapman2006,Jahanbakhsh2017}. Yet, they can also reinforce users' biases and preferences to work with familiar and similar individuals \cite{Gomez-Zara2019-gg, Piezunka2015-rf, Hinds2000-uu}. Assessing potential collaborators is also difficult because skills, points of view, and knowledge characteristics are intangible, leading people to judge others based on first impressions, stereotypes, social references, and demographics \cite{Contractor2013-so}.

A third possibility is to incorporate organizational values into users' searches and choices without removing their agency. For instance, algorithms can nudge choices by increasing the visibility of collaborators who may be more suitable for the organization's goals \cite{Liu2023,zheng2017hybrid,Zhang2024}. Algorithmic nudging refers to a subtle form of modifying choice architectures to guide user behavior in digital environments without restricting the individual's freedom of choice, often operating without users' full awareness \cite{weinmann2016digital,jesse2021digital,yeung2019hypernudge}. This approach can preserve user choice while subtly steering decisions toward organizational and collective goals \cite{kamar2012combining}. For instance, LinkedIn researchers tested a re-ranking approach that adjusted top-ranked candidate lists to better reflect the gender distribution of the qualified candidate pool, producing a nearly threefold increase in search queries with representative results \cite{Geyik2019-qt}. These ``algorithmic nudges'' have profound implications for team composition, collective decision-making, and organizational values \cite{Kellogg2020-oo, Polzer2022-pl,Raveendhran2021-lc}. 

Despite the growing deployment of these algorithms across collaboration and social platforms, we lack systematic evidence on how different team formation approaches compare in their effects on collaborator selection, team composition, team dynamics, and outcomes. Assignment, free choice, and nudging each embed different assumptions about autonomy, transparency, and organizational control \cite{Kellogg2020-oo,Lee2015}, and each can produce unintended consequences even when designed with benevolent purposes \cite{Selbst2019}. Assignment can efficiently achieve organizational goals, but it removes individual choice, potentially triggering resistance and disengagement \cite{Dobbin2016-xt,Dover2016-xx,kalev2006best}. Free choice respects autonomy but reinforces homophily---the tendency to prefer similar others \cite{McPherson2001-ei,Ertug2022}---which may narrow collaboration opportunities for individuals with fewer existing social ties \cite{Ertug2022,Hinds2000-uu}. Nudging preserves user agency while subtly steering decisions, but it can operate with limited transparency and shape who gains access to collaborators through mechanisms users do not fully perceive \cite{botes2023autonomy,yeung2019hypernudge}. More fundamentally, we do not yet understand how these algorithmic designs shape individuals' opportunities to enter and build collaboration networks \cite{Raghavan2020,Fabris2025}. This study addresses this gap by asking the following research question: \textit{How do variations in user agency and organizational values in different team formation algorithms shape individuals' collaboration opportunities, team composition, and collective outcomes?}

To address this research question, we conducted a between-subjects laboratory experiment investigating the impact of recommendation system algorithms on individuals' collaborator choices when forming groups. Employing a $2 \times 2$ factorial design, we designed four team formation algorithms that manipulated \textit{user agency} (i.e., assignment vs. choice) and \textit{heterogeneity criteria} (included vs. not included), resulting in four conditions: random assignment, assigned-diverse assignment, self-assembled teams (i.e., users searching for collaborators and forming teams on the recommendation system), and nudged teams (i.e., re-ranking collaborator recommendations based on heterogeneity criteria). We deployed these algorithms in an existing team recommendation platform \cite{contractor2013my} and recruited over 330 participants, resulting in 83 teams in total. Participants were randomly assigned to one of these algorithms---unaware of which condition they were in until the experiment concluded---and the resulting teams worked on a creativity task requiring them to ideate, generate, discuss, and implement ideas. We analyzed how these algorithms influenced collaborator selection, team composition, team dynamics, and team outcomes through survey responses, trace data, conversation transcripts, and external evaluations of team deliverables. The findings demonstrate that algorithms can influence individuals' collaboration choices and improve team outcomes, yet their deployment requires careful consideration of the tension between organizational goals and individual autonomy, as these interventions operate largely outside users' awareness and control. 


Our contributions are threefold. First, this study extends insights into algorithmic management by providing causal evidence on how two key algorithmic design choices (user agency and heterogeneity criteria) independently and jointly shape team formation from initial selections through to team outcomes. Second, we present a comprehensive laboratory experiment showing that the effects of organizational values---in this case, heterogeneity criteria---depend not only on \textit{whether} they are included, but also on \textit{how} they are incorporated into algorithms. Lastly, we contribute to emerging scholarship on algorithmic governance \cite{Martin2019-dd,Kellogg2020-oo} by demonstrating that algorithms embedded in social/collaboration platforms constitute a distinct mode of organizational influence, one that operates upstream of work execution by shaping social opportunities rather than directly controlling tasks or performance. We discuss the ethical implications of algorithmic nudging for team formation and considerations for socio-technical systems that strike a balance between effectiveness, user autonomy, and transparency.

%% file: 02_Literature_Review.tex
\section{Literature Review}
\label{sec:literature_review}

\subsection{Algorithmic Management and Governance in Organizations}
Prior research conceptualizes algorithms as computational systems that transform input data into decisions that assign, evaluate, and coordinate work \cite{Kellogg2020-oo}. In organizational contexts, algorithms exert control through mechanisms such as restricting options (e.g., assignment) or recommending alternatives (e.g., ranked suggestions), both of which are central to team formation. Importantly, these systems embed organizational values and ethical assumptions through their design choices, data sources, and optimization goals \cite{Kitchin2017-wy,Martin2019-dd}.

Research on algorithmic management has largely focused on how algorithms direct or evaluate individual workers, particularly in gig and platform labor. This work shows that algorithmic control can reduce perceived autonomy and fairness, even when formal choice remains \cite{Lee2015,benlian2022algorithmic,cram2022examining}. Perceptions of fairness depend on how decisions are communicated and whether workers perceive opportunities for recourse \cite{lee2018understanding,Sannon2022}. However, comparatively little attention has been paid to how algorithms shape collaborative decisions—specifically, how individuals find, select, and form teams with others. This gap is consequential as collaboration increasingly occurs through digital platforms that mediate access to collaborators and opportunities \cite{Faraj2018-tz,Polzer2022-pl}.

Recent scholarship frames algorithms as a mode of governance that shapes decision environments rather than issuing direct commands \cite{yeung2019hypernudge,parent2022algorithms}. This perspective foregrounds questions of legitimacy, accountability, and contestability in algorithmic systems \cite{Selbst2019,Vaccaro2020,alfrink2023contestable}. These concerns are especially salient for team formation, where algorithmic decisions structure professional networks and long-term opportunities \cite{Gomez-Zara-2020-cscw}.

\subsection{User Agency in Team Formation}
User agency---the capacity to exercise control over one’s decisions and actions \cite{Bandura1986-da}---is increasingly mediated by algorithmic systems. While algorithms can support decision-making through search and recommendation, they can also constrain agency through design choices that shape available options and salience \cite{Holton2021,Lyons2021}. Recommendation systems, widely used across domains, exemplify this tension: they preserve formal choice while influencing behavior through ranking and exposure effects \cite{Chen2013-cl,Jameson2015-lv,Geyik2018-zx}.

In team formation, a core design choice concerns whether users retain agency over collaborator selection or are assigned to teams algorithmically \cite{Gomez-Zara-2020-cscw}. Assignment-based approaches enable organizations to directly optimize for criteria such as skill complementarity or heterogeneity \cite{Lykourentzou2017-ui,Barnabo2019}, and have been shown to support coordination among strangers \cite{Valentine2017-pp,Salehi2017}. However, removing choice can reduce engagement and commitment, particularly when team members are unfamiliar or perceived as externally imposed \cite{Chapman2006,Dobbin2016-xt}.

Conversely, self-assembled teams benefit from members' motivation, familiarity, and sense of ownership \cite{Contractor2013-so,wax2017self}. Yet free choice is subject to homophily, leading individuals to prefer similar collaborators and perpetuating unequal access to collaboration opportunities \cite{McPherson2001-ei,Gomez-Zara2019-gg,kleinbaum2013discretion}. This tension motivates exploration of intermediate approaches that preserve agency while shaping collaboration patterns.

\subsection{Algorithmic Nudging and Choice Architecture}
Between full assignment and unconstrained choice lies a third possibility: algorithmic nudging. Drawing on behavioral economics, nudging refers to interventions that guide behavior by modifying the choice architecture (i.e., the environment in which decisions are made) without restricting options or significantly altering economic incentives \cite{thaler2021nudge}. Nudges work by leveraging cognitive heuristics and biases, such as defaults, salience, and social proof, to steer individuals toward outcomes deemed beneficial by the choice architect.

In digital environments, nudging takes the form of default settings, personalized recommendations, and interface design choices that make certain options more salient or accessible \cite{weinmann2016digital}. Digital nudges can be particularly powerful because they can be personalized at scale and adapted in real time based on user behavior \cite{Yeung02012017}. Weinmann et al. \cite{weinmann2016digital} define digital nudging as ``the use of user-interface design elements to guide people's behavior in digital choice environments.'' Jesse and Jannach \cite{jesse2021digital} provide a comprehensive survey of nudging in recommender systems, identifying common techniques including default recommendations, social proof signals, and strategic ordering of options. Caraban et al. \cite{caraban201923} catalogued 23 distinct nudging mechanisms used in human-computer interaction, ranging from subtle interface changes to more explicit prompts. Research on AI-assisted decision making has characterized the cognitive mechanisms through which algorithmic nudges operate. Rastogi et al. \cite{Rastogi2022} found that AI recommendations can induce anchoring bias, leading users to insufficiently adjust away from these recommendations, and demonstrated that allocating additional cognitive resources (e.g., time) can mitigate this bias. These findings highlight that nudging operates not merely through information provision but through the shaping of cognitive processes and attention.

Nudging raises ethical questions that are particularly salient when applied to organizational decisions affecting individuals' opportunities. Unlike transparent policies, nudges often operate without users' full awareness since they shape decisions through mechanisms users do not perceive \cite{Eslami2015,jesse2021digital}. For instance, Eslami et al. \cite{Eslami2015} found that most Facebook users were unaware of algorithmic curation of their news feeds, and when informed, expressed surprise and concern. Furthermore, nudges designed to benefit organizations may not align with the interests of individual users. Holstein et al. \cite{Holstein2019} found that industry practitioners struggle to reconcile fairness goals with other system objectives, while Madaio et al. \cite{Madaio2020} documented organizational challenges in implementing fair AI systems. Kawakami et al. \cite{Kawakami2022} studied AI-assisted decision making in child welfare, finding that workers expressed ambivalence about algorithmic recommendations that conflicted with their professional judgment. These tensions are compounded by the fact that nudges, like other algorithmic interventions, can have unintended consequences, potentially advantaging some individuals while disadvantaging others \cite{Selbst2019,De-Arteaga2019-ah}.

In the context of team formation, nudging can take the form of reordering collaborator recommendations to increase the visibility of candidates who would add variation to the team. Prior work shows that re-ranking can improve representational balance without restricting choice \cite{Geyik2019-qt}, though explicitly signaling candidates' differences can produce complex or counterproductive effects \cite{Gomez-Zara2020-ol}. Building on this literature, our study compares nudging to assignment and free choice to examine how different mechanisms for shaping collaboration affect team composition, processes, and outcomes.

\subsection{Team Heterogeneity}
Algorithms may also be used to shape team composition by promoting heterogeneity among team members. However, heterogeneity is not a unitary construct. It spans multiple dimensions---such as demographic attributes, professional backgrounds, and perspectives---that differ in how they are operationalized, perceived by team members, and linked to collaboration outcomes \cite{Harrison2007-fl,qin2014review}. Researchers distinguish between \textit{surface-level} and \textit{deep-level} differences, which have different antecedents, are affected differently by team formation mechanisms, and have distinct implications for team processes and outcomes \cite{Harrison1998-hs,Harrison2002-xd,Page2019-dm}.

Surface-level differences refer to variations in readily observable attributes, such as gender, age, and race \cite{Harrison1998-hs}. These attributes are overt, immediately visible to team members and to individuals evaluating potential collaborations. Promoting surface-level differences in teams can bring different traits, perspectives, and experiences inherent to the demographic group \cite{Van_Knippenberg2004-ia,Page2008}. Some examples include gender heterogeneity, which promotes productivity in software development teams \cite{Vasilescu2015-nc} and collective intelligence within teams \cite{woolley2010evidence}. However, because surface-level attributes are salient during initial interactions, they tend to activate social categorization processes (i.e., ``us-them'' distinction), potentially triggering intergroup biases that negatively affect communication, cohesion, and coordination \cite{Williams1998-rr,Tajfel1979-yu}. 

Deep-level differences refer to variations in less observable attributes, such as knowledge, skills, values, attitudes, and cognitive styles \cite{Harrison1998-hs}, which ultimately become apparent through extended interaction. Deep-level heterogeneity is often considered more relevant to team performance since it reflects the variety of information, perspectives, and approaches that team members bring to tasks \cite{Van_Knippenberg2004-ia}. For instance, teams with diverse knowledge and skills can draw on a broader range of resources when solving problems, potentially leading to more creative and higher-quality outcomes \cite{Bell2011-wu,Horwitz2007-fu,Uzzi2013-ap}. However, it can also create coordination challenges and disagreements about how to approach tasks \cite{Jehn1999}.  

Research on team heterogeneity has produced mixed findings, with meta-analyses suggesting that the effects depend on the type of heterogeneity, the nature of the task, and the management approach \cite{Bell2011-wu, Horwitz2007-fu}. Moreover, heterogeneity can lead to coordination problems and conflicts within a group due to differences in training, values, and knowledge \cite{Harrison1998-hs,Lau2005-pb}. Previous studies have also found no direct effects of heterogeneity on performance \cite{Stewart2020,Bell2011-wu}. Some researchers have argued that how participants perceived their differences could be a better predictor than the factual team composition \cite{Shemla2016-vs}. These nuances have led scholars to characterize team heterogeneity as a ``double-edged sword'' \cite{Horwitz2007-fu}, offering informational benefits while potentially triggering costs within groups \cite{Van_Knippenberg2004-ia, OReilly1998-jd}. Overall, the interplay between surface-level and deep-level differences plays a role in how team members' differences influence their dynamics and outcomes.

Algorithmic approaches to heterogeneity further complicate these dynamics. Scholars argue that heterogeneity is a contested value in computational systems, and that different operationalizations produce different distributional consequences \cite{Fazelpour2022,De-Arteaga2019-ah}. Despite extensive research on heterogeneity outcomes, few studies examine how team formation mechanisms shape both the level and type of heterogeneity that emerges, as well as how these mechanisms influence team processes. This study addresses this gap by comparing assignment, nudging, and free choice in algorithmic team formation.

\subsection{Hypotheses}
Building on the preceding review, we develop hypotheses about how manipulating user agency and heterogeneity criteria in team formation algorithms influence collaborator selection, team composition, and team outcomes. We also pose a research question about team processes, which we examine in an exploratory manner.

\subsubsection{Collaborator Selection}
When individuals search for collaborators on a recommendation system, their selections will be influenced by both their preferences and the options presented by the algorithm. Recommendation systems shape user behavior by determining which options are most visible and accessible \cite{jannach2010recommender,jesse2021digital}. Due to position effects, users are likely to disproportionately attend to top-ranked options, and higher-ranked items receive more clicks, regardless of intrinsic fit with their preferences \cite{Geyik2018-zx,chun2022power}. In team formation contexts, this suggests that collaborators appearing at the top of search results will be selected more frequently, independent of user preferences.

When algorithms incorporate heterogeneity criteria by reordering recommendations---placing collaborators who would increase team heterogeneity in more prominent positions---users should be more likely to encounter and select such collaborators. This nudging effect operates through increased visibility rather than restricting choice: users can still select anyone, but options that would add compositional variation become more salient. Research on digital nudging suggests that such visibility interventions can significantly shift behavior without removing choice \cite{weinmann2016digital, caraban201923}. 

We expect nudging effects to be particularly pronounced for attributes that are easily inferable from the recommendation interface. In our system, participants viewed the names and profile photos of potential collaborators. Names often signal gender \cite{bayerl2024gender} and can indicate racial or ethnic background \cite{bertrand2005implicit}, while profile photos make both race and gender visually apparent \cite{Jahanbakhsh2020}. Prior research documents that such demographic cues activate homophily preferences, leading individuals to favor demographically similar collaborators \cite{McPherson2001-ei}. By reordering recommendations to increase the visibility of collaborators who would increase demographic heterogeneity, the algorithm may counteract these tendencies. We hypothesize that a higher ranking serves as a validation mechanism, signaling that collaborators from different demographic backgrounds are suitable choices. Because both race and gender were visible in our interface and both are subject to homophily, we test nudging effects for each attribute separately. Thus, our first set of hypotheses is:

\begin{quote}
    \textbf{H1a (Recommendation Effect)}: \textit{Users will be more likely to select collaborators who rank higher in the algorithm's recommendations.}\\
    \textbf{H1b (Nudging Effect on Different-Race Collaborator Selection)}: \textit{When a recommendation algorithm incorporates racial heterogeneity criteria, it will nudge users toward selecting collaborators from different racial groups than their own.}\\
    \textbf{H1c (Nudging Effect on Different-Gender Collaborator Selection)}: \textit{When a recommendation algorithm incorporates gender heterogeneity criteria, it will nudge users toward selecting collaborators from different gender groups than their own.}
\end{quote}

\subsubsection{Agency Effects on Team Composition}
The composition of teams depends on both the algorithm's design and users' responses to it. When algorithms remove user agency and assign individuals to teams, they can optimize directly for heterogeneity criteria because users cannot override team assignments through their own collaborator preferences. In contrast, when algorithms allow users to choose their own collaborators, homophily will likely shape the resulting composition. In team formation contexts, this manifests as individuals' preferences for collaborators who share demographic characteristics, shared attributes, social familiarity, or similar backgrounds \cite{McPherson2001-ei,Gomez-Zara2019-gg}. We therefore expect:

\begin{quote}
    \textbf{H2a (Agency Effect on Surface-Level Differences)}: \textit{Teams formed using algorithms that allow user agency will exhibit lower surface-level differences than teams formed through algorithmic assignment.}\\
    \textbf{H2b (Agency Effect on Deep-Level Differences)}: \textit{Teams formed using algorithms that allow user agency will exhibit lower deep-level differences than teams formed through algorithmic assignment.}
\end{quote}

\subsubsection{Heterogeneity Criteria Effects on Team Composition}
Algorithms that incorporate heterogeneity criteria---whether through assignment or nudging---should produce more heterogeneous teams than those that do not. In the assignment condition, heterogeneity criteria directly determine team composition through an optimization process. In the agency condition, heterogeneity criteria increase the visibility of collaborators who would add compositional variation, shifting selections toward greater team heterogeneity even if not to the same degree as assignment. Research on ranking has demonstrated that reordering recommendations can significantly affect outcome distributions without restricting choice \cite{Geyik2019-qt,singh2018fairness,biega2018equity}. We therefore expect:

\begin{quote}
    \textbf{H3a (Heterogeneity Criteria Effect on Surface-Level Differences)}: \textit{Teams formed using algorithms that incorporate heterogeneity criteria will exhibit higher surface-level differences than teams formed without heterogeneity criteria.} \\
    \textbf{H3b (Heterogeneity Criteria Effect on Deep-Level Differences)}: \textit{Teams formed using algorithms that incorporate heterogeneity criteria will exhibit higher deep-level differences than teams formed without heterogeneity criteria.}
\end{quote}

\subsubsection{Team Processes}
Team formation does not end when teams are assembled. How teams are formed may shape members' initial perceptions and interaction patterns, with downstream consequences for collaboration quality \cite{tuckman1977stages}. We develop two specific hypotheses about process variables central to our theoretical argument. First, we expect that the mechanism by which heterogeneity is achieved will influence how salient member differences become. When teams are assigned to maximize heterogeneity, members may experience their team composition as externally imposed and as a constraint on their autonomy, heightening attention to surface-level and deep-level differences \cite{Dobbin2016-xt}. In contrast, when members choose their collaborators---even if nudged toward a more varied composition---the team may feel greater self-determination and commitment to the team, reducing the salience of demographic and cognitive differences \cite{Shemla2016-vs,Chapman2006}. We therefore expect:

\begin{quote}
    \textbf{H4a (Perceived Surface-Level Differences)}: \textit{Members of teams formed through assignment will report higher awareness of surface-level differences than members of teams formed through algorithms that allow user agency.} \\
    \textbf{H4b (Perceived Deep-Level Differences)}: \textit{Members of teams formed through assignment will report higher awareness of deep-level differences than members of teams formed through algorithms that allow user agency.}
\end{quote}

Second, we expect that how team heterogeneity is achieved will affect subsequent communication patterns. When heterogeneous teams are formed through algorithmic assignment, members must build relationships with unfamiliar individuals they did not choose. This situation may create intergroup anxiety and hesitation that inhibits communication \cite{Stewart2020,Tajfel1979-yu}, particularly when demographic differences are salient \cite{Bell2011-wu}. In contrast, when team heterogeneity emerges through user choice---even if nudged by algorithmic recommendations---members enter the collaboration with a stronger sense of ownership and mutual interest that may buffer against these communication barriers \cite{Chapman2006}. Because they played a role in forming the team, they may be more likely to initiate conversations and coordinate with one another. We therefore expect:

\begin{quote}
    \textbf{H5 (Communication Frequency):} \textit{Members of assigned teams will report less frequent communication with teammates than members of teams formed through algorithms that allow user agency.}
\end{quote}

\subsubsection{Team Outcomes}
Finally, we examine how the combination of user agency and heterogeneity criteria affects team outcomes. We expect nudged teams to achieve the best outcomes. This prediction rests on two arguments. First, nudging could support a form of balanced heterogeneity. Prior research on team composition suggests that variation in members' knowledge, skills, and perspectives can improve team outcomes when teams exchange and integrate those differences \cite{Van_Knippenberg2004-ia,Bell2011-wu}. These benefits are more likely to emerge when members feel comfortable and committed enough to engage with one another across differences \cite{Dobbin2016-xt,Shemla2016-vs}. Assignment can directly create compositionally varied teams, but it may undermine the comfort and engagement needed to use those differences effectively. Free choice can preserve commitment, but it may produce homogeneous teams that lack distinct perspectives. Nudging offers a middle ground, as users can choose their teammates from a subtly reordered choice set that increases exposure to different collaborators while preserving agency and ownership.

Second, research on choice architecture suggests that guided choice can improve decisions by structuring options without removing agency. When individuals make decisions in well-designed choice environments, guidance can reduce biased selections while preserving engagement \cite{Heer2019,thaler2021nudge}. In team formation, this suggests that nudging may combine sufficient heterogeneity with sufficient member commitment to enable effective interaction and collaboration. Therefore, our last hypothesis is:

\begin{quote}
    \textbf{H6 (Nudging Effect on Outcomes):} \textit{Teams formed using algorithms that combine user agency with heterogeneity criteria will achieve better outcomes than teams formed using algorithms with only agency, only heterogeneity criteria, or neither.}    
\end{quote}

%% file: 03_Methodology.tex
\section{Methodology}
\label{sec:methodology}
We conducted a between-subjects laboratory experiment with a 2 $\times$ 2 design implemented using a team formation recommendation system called ``My Dream Team'' \cite{contractor2013my}. We implemented four algorithms on this system that manipulated user agency (included vs. not included) and heterogeneity criteria (included vs. not included). Depending on the condition, participants either selected collaborators through the recommendation system or were assigned to teams by the system, and then worked as members of the resulting teams. We then collected individual-level data from participants and team-level data from the resulting teams to test our hypotheses. We pre-registered the study before conducting any data analysis \cite{gomez_zara_2022_osf_registration}. \anon{Northwestern University}'s Institutional Review Board approved this study (IRB Protocol Number \# \anon{STU00212865}). Data collection occurred between July and November 2021. 

\subsection{Sample Size}
We used \textit{G*Power 3.1} \cite{faul2009statistical} to estimate the required sample size for our experiment. We specified an F-test for fixed-effects ANOVA using the main-effects-and-interactions option, with four groups, numerator $df=1$, an assumed large effect size of Cohen's $f = 0.4$, a significance level of $\alpha=0.05$, and power of 0.95 ($1 - \beta$). This calculation yielded a target sample size of 84 teams. 

\subsection{Participants}
We recruited participants through \anon{Northwestern Kellogg}'s Behavioral Research Lab and prescreened them to verify their qualifications for this study. The requirements were being 18 or older, residing in the United States, and being a full-time employee or student. The participant pool consisted of the institution's students, staff, faculty, and local area residents. More than 600 participants answered a battery of demographic, educational, and social network questionnaires. The goal was to conduct experimental sessions using a specific team formation algorithm.   

To facilitate the experiment's execution and accommodate participants' availability, we conducted experimental sessions involving 30 to 50 participants at a time. This decision was made to effectively manage logistical constraints, such as coordinating multiple participants and mitigating potential technical challenges that could arise with larger groups. Moreover, medium sessions enabled us to maintain consistency in team formation and minimize delays, ensuring equitable participation for all individuals. We used stratified sampling to ensure a balanced demographic distribution in each experimental session.

After prescreening and session assignment, we invited approximately 500 qualified subjects to participate in this study via email. The invitation email contained a brief introduction to the study, the payment information, and a confirmation link to the electronic consent form. The final sample consisted of 386 participants who agreed to participate in the study. From this sample, 197 participants were students, and 189 were community members (46 university employees and 143 local residents). Regarding gender, 124 participants reported being male, 252 female, and 10 non-binary. The sample consisted of 194 White, 123 Asian, 41 African American, and one American Indian participants. Twenty participants identified as belonging to multiple races. Regarding their country of origin, 314 identified themselves as from the US and 72 as from another country. Lastly, 35 participants were Hispanic or Latino. 

We compensated each participant \$40 USD; each session took about 120 minutes. We also announced that the top-performing team of this study will receive an additional \$50 USD gift card per member. Due to COVID-19 pandemic restrictions, we conducted these sessions online by hosting meetings on Zoom. These virtual meetings enabled participants to use their voices and see each other's faces, emulating face-to-face interactions. We did not control participants' location due to COVID-19 pandemic lockdowns. 

\subsection{Team Task}
Each team completed a 30-minute creativity task adapted from Klonek et al. \cite{Klonek2021-gv}, which required developing volunteer recruitment materials for a fictitious healthcare organization. The campaign’s goal was to recruit volunteer drivers to transport cancer patients from their homes to hospitals. We selected a creativity task for three reasons. First, following McGrath’s circumplex \cite{mcgrath1984groups}, creativity tasks are open-ended, lack unique correct answers, benefit from various perspectives, and require coordination to produce a shared deliverable, making them well-suited for observing early-stage collaboration dynamics. Second, creativity tasks are theoretically relevant for studying team composition because heterogeneous teams are expected to outperform homogeneous teams when tasks require integrating multiple perspectives and generating novel ideas \cite{Page2019-dm,Van_Knippenberg2004-ia}. Third, similar creativity-based tasks have been widely used in CSCW and team research to study collaboration processes, including brainstorming \cite{Dow2011,Shi2017}, ideation \cite{Whiting2020,salehi2018}, and design work \cite{Lykourentzou2017-ui,valentine2025flash}. All participants received background information about the organization and its need for volunteers.

\subsection{Team Size}
Because team heterogeneity depends in part on team size, we held team size constant across conditions. We set the target team size to four members, which ensured sufficient within-team variation to compute team composition measures while keeping teams comparable. When the number of participants in a session was not divisible by four, some participants were unable to assemble a team and completed the task as solo participants. We excluded them from our data analysis.

\subsection{Procedure}
\begin{figure}[!htb]
    \centering
    \includegraphics[width=0.9\linewidth]{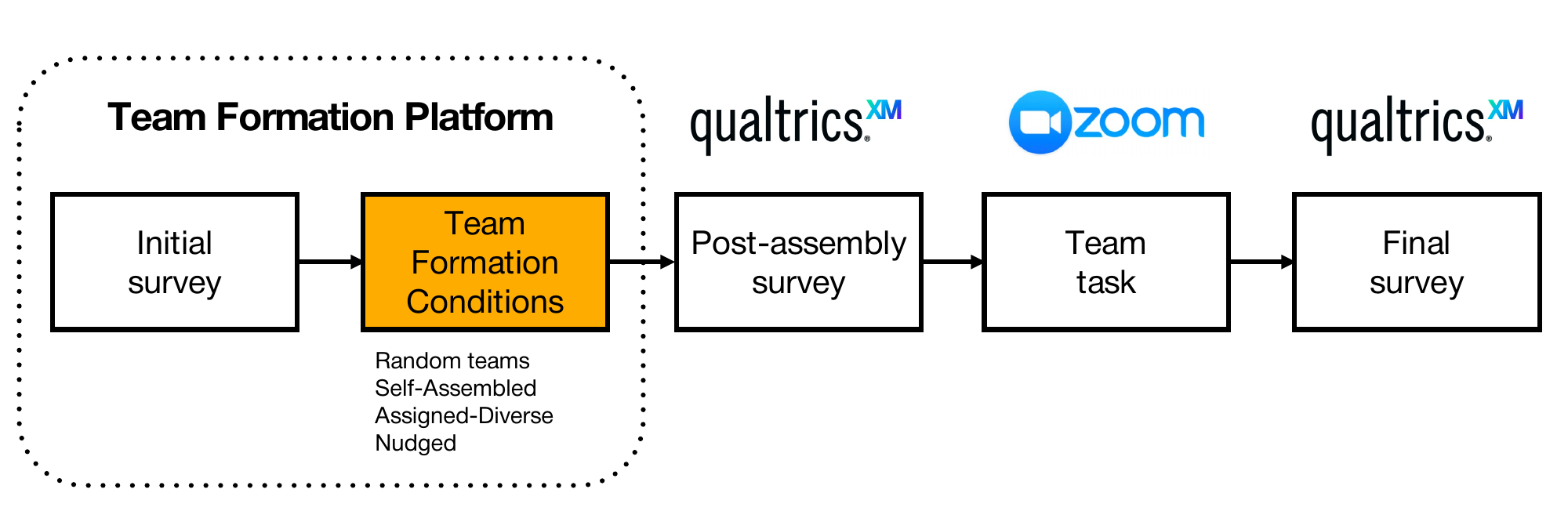}
    \Description{This diagram describes how the research team conducted the data collection process}
    \caption{Study outline. This diagram describes how the research team conducted the data collection process.}
    \label{fig:study-outline}
\end{figure}

Figure \ref{fig:study-outline} presents the main study outline. For each experimental session, we invited participants to Zoom meetings and asked them to keep their cameras off and microphones muted. The research team developed the `My Dream Team' recommender system to facilitate data collection and conduct the proposed experimental conditions. The system was implemented using \textit{Django} and \textit{Python}. We created accounts for each participant and sent the login instructions to their email addresses provided during registration. 

\subsubsection{Initial Survey}
At the beginning of each experimental session, we described the study's purpose and asked participants to populate a profile on the recommendation system. They completed a survey to provide demographic information, personality traits, leadership, creativity, psychological collectivism, and social skills, which fed the information necessary to create the system's teammate recommendations. We asked them to identify other participants they knew, had previously collaborated with, and enjoyed socializing with. In addition, the survey included a set of self-assessment questions related to the team task, measured on a 5-point Likert scale ranging from ``Not at all skilled'' to ``Extremely skilled.'' These items assessed participants’ perceived skills in managing campaigns, coordinating people, designing visual graphics, recruiting volunteers, writing reports, articles, or papers, and presenting reports and new ideas. All responses were kept confidential and were used by the recommendation system to compute the recommended teams. After all participants had created their profiles and filled out the initial survey, we introduced them to the team task and prepared them for the team formation stage. 

\subsubsection{Team Formation Conditions}
\label{team_conditions}
After explaining the task to the participants, we randomly assigned sessions to one of the following experimental conditions within the $2 \times 2$ experimental design. 

\paragraph{Random teams} (i.e., no heterogeneity criteria, no agency): The team formation recommendation system assigned participants to teams randomly. Participants saw the teams assigned to them by the recommendation system. If the total number of participants was not a multiple of four, the system randomly assigned some participants as solos.

\paragraph{Self-Assembled teams} (i.e., no heterogeneity criteria, agency included): The recommendation system prompted the participants with a search interface to find potential collaborators. This interface consisted of a search query to indicate collaborator preferences based on the attributes collected in the initial survey. Some search criteria examples included finding people with high expertise, similar or different characteristics, or people with creative skills. In each search query, participants had to select at least two criteria and rate the importance of these criteria using a 7-point Likert scale, ranging from ``Not important at all'' (-3), to ``Don't care'' (0), to ``Yes, for sure.'' (+3). By default, all the criteria were set to zero, and the participant had to manually adjust their importance to create a query. The system computed a ``fit score'' for each potential collaborator $j$ matching the searcher participant $i$'s queries according to the following formula: $S_{i,j}= \sum_{k \in K} \alpha_k s_{i,j,k}$, where $k$ is the importance of the criteria $k$ in the search query. Then, each fit score $s_{i,j,k}$ considers the potential collaborator $j$'s score of the criterion $k$ in relation to the searcher $i$. The recommendation system used the $S_{i,j}$ scores to display a rank-ordered list of the potential collaborators that best fit the participant $i$'s query. 

The system paginated all the results ordered according to the participant's query. For each collaborator recommendation, the system displayed the participants' full name, the percentage of how well they matched with the participant's query, a link to their public profile, their current teammates, and an \textit{``Invite''} button (see Appendix Figure \ref{fig:recommender-system}). If the participant (sender) invites the potential collaborator (recipient), the system sends an invitation message to that person, their teammates, and the sender's team. Participants who received an invitation could accept it, decline it, or ignore it. If all recipients accepted the invitation, the sender and the recipients would be in a group. If the recipients or sender were already part of a group, their groups would merge. The recommendation system allowed groups to merge only if the final group size was equal to or less than four. Participants were also able to switch or leave their groups. Participants could also view the list of assembled teams and invite members from there. Participants had ten minutes to assemble their groups. Those who were not part of a team of four were either assigned to work individually or placed in teams with fewer members.

\paragraph{Assigned-Diverse teams} (i.e., heterogeneity criteria included, agency not included): The recommendation system assigned participants to teams using a multi-objective genetic algorithm that maximized teams' surface-level differences (i.e., participants' age, gender, and race) and deep-level differences based on the project skills required to solve the experiment's task \cite{Gomez-Zara2022-av}. This algorithm started by creating random team combinations and then searched for better combinations by swapping members. If the resulting team combination had higher heterogeneity scores than the previous ones, the new combination was kept, and the previous ones were deleted. This process was repeated twenty times, creating a Pareto front that maximized both surface-level and deep-level heterogeneity scores. We employed the elbow method to select one team combination from the Pareto Front and assign participants to teams. Participants saw the teams assigned to them by the recommendation system after the task explanation. If the total number of participants was not a multiple of four, the algorithm assigned some participants as solos during the optimization process.

\paragraph{Nudged teams} (i.e., heterogeneity criteria and agency included): As in the Self-Assembled teams' condition, the recommendation system prompted participants with a search interface to find collaborators. In this condition, however, the recommendation algorithm re-ranked the teammate recommendations using heterogeneity criteria. The system calculated a ``heterogeneity score'' $H_{i,j}$ for each potential teammate $j$, estimating how much the recommended teammate $j$ would change the composition of the searcher $i$'s current team (metrics described in Section \ref{subsec:measures}). The algorithm multiplied the fit score ($S_{i,j}$) by the heterogeneity score ($H_{i,j}$) and used the resulting score to re-rank recommendations. This approach kept recommendations aligned with searchers' queries while also prioritizing teammates who would increase team heterogeneity. Rather than optimizing for a single attribute, the algorithm weighted all composition metrics equally to balance them comprehensively. As a result, participants who contributed less to team heterogeneity appeared lower in the ranking, while those who contributed more ranked higher. Participants had ten minutes to assemble their teams. Participants who were not part of a team of four were either in teams with fewer members or assigned to work as solos. 

\subsubsection{Post-Assembly Survey}
Once the teams were assembled, participants had three minutes to review their teammates' profiles on the recommendation system, which included their names and public profiles. Participants then completed a post-assembly survey to evaluate the recommendation system and assess their experience in forming teams with it (Appendix Table \ref{tab:post-assembly}). The survey was conducted on \textit{Qualtrics}, and we sent a personalized link to each participant.

We included manipulation check questions to assess whether participants perceived the experimental interventions. Using 5-point Likert scales, we asked to what extent they could choose their collaborations (i.e., agency) and whether their teams were heterogeneous (i.e., heterogeneity criteria). We employed linear regressions to estimate the values of these questions using participants' answers based on the respective manipulation variables. Overall, participants in the agency conditions were aware of whether they could choose teammates ($p < 0.001$). Participants in the heterogeneity criteria conditions were not significantly aware that their teams were heterogeneous ($p = 0.09$), possibly because they had not interacted with their collaborators before answering this question.

\subsubsection{Team Task}
Participants were then moved into Zoom breakout rooms with their assembled teams. They had to activate their cameras and microphones in their breakout rooms. Each breakout room had an assigned computer for the research team to record their discussion. Each team received a Google Slides deck to create their recruitment materials in the breakout room's chat. Team participants were encouraged to discuss and solve the problem as a group. We gave teams 30 minutes to complete the task. Alerts were sent to all breakout rooms five minutes before the end of the task. We also moved solo participants to breakout rooms so that they could work alone. 

\subsubsection{Final Survey}
After participants completed their team task, we closed the Zoom breakout rooms and asked them to turn off their cameras and microphones. We sent participants a personal link to a final survey on Qualtrics. This final survey aimed to understand participants' perceptions of the various team processes, dynamics, and experiences working with their teammates. We also included open-ended questions to capture participants' experiences with the recommendation system when assembling their teams. On average, this survey required 20 minutes. 

After completing the experiment, participants were fully debriefed about the study’s purpose and informed that the team formation system operated under one of four experimental conditions. We explained that some details of the recommendation process were withheld prior to participation to preserve experimental validity, in line with established A/B testing practices in online behavioral research \cite{kohavi2009controlled,kohavi2020trustworthy}. Participants were informed of their specific condition, the rationale for this incomplete disclosure, and assured that the intervention posed minimal risk and did not affect their rights or welfare. Participants received their compensation after their session was completed. We awarded a bonus to the team with the highest creativity score after data collection was completed.

\subsection{Analysis}
\label{subsec:measures}
Our analysis proceeds in four stages corresponding to our hypotheses: collaborator selections, team composition, team processes, and team outcomes. These analyses operate at different levels. Collaborator selection was analyzed at the individual level, with each algorithm-generated recommendation as an observation ($N = 4,351$ recommendations from 185 participants). Team composition, outcomes, and transcript analyses were conducted at the team level ($N = 83$ teams). Team process variables were measured at the individual level ($N = 332$ participants), with each participant reporting their perceptions of team member differences and communication.

\subsubsection{Collaborator Selections}
To test H1a-c, we analyzed trace data from the recommendation system in the agency conditions, treating participants' invitations as the dependent variable. We employed generalized logistic mixed-effects regression models to assess which factors were likely to explain these collaborator choices. We included random effects for participants as they could have generated multiple searches. The model's observations were the recommendations generated by the agency algorithms, and the dependent variable was whether a recommended collaborator was invited by the searcher. 

For the fixed effects, we included the fit score described in Section \ref{team_conditions} to test whether higher-ranked recommendations increased the likelihood of selection (H1a). We also added a ``different gender'' variable to assess gender homophily in collaborator choices. Similarly, we added a ``different-race'' variable to assess race homophily in their choices. We treated the inclusion of heterogeneity criteria as a treatment variable: 0 in the Self-Assembled condition (the algorithm used only the fit score to sort the recommendations) and 1 in the Nudged condition (the algorithm used both fit and heterogeneity scores). To test H1b, we included an interaction term between the treatment and different-gender variables, and to test H1c, an interaction term between the treatment and different-race variables. These terms allowed us to examine whether participants in the Nudged condition were more likely to select different collaborators. 

We computed a GLME model using the binomial link function, given the non-normal distribution of the data. We scaled variables before analysis and verified that multicollinearity was not an issue by checking that the Variance Inflation Factor (VIF) scores for the independent variables were all below 10. We conducted the statistical tests using \textit{R 4.4.0} and using the \texttt{lme4} package.

\subsubsection{Team Composition}
To test our composition hypotheses, we operationalized team heterogeneity as variation in members' demographic and task-relevant attributes, following Harrison and Klein's typology \cite{Harrison2007-fl}. For categorical attributes collected in the initial survey (i.e., gender, race, international/domestic, and ethnicity), we calculated team heterogeneity using the Blau index as $1-\sum p_{i}^2$, where $p_i$ is the proportion of team members in category $i$. A high score indicates that members fall into different categories for that attribute, while a low score means that all members fall into the same category. For numerical attributes (i.e., participants' age and their six project skills), we calculated heterogeneity using the coefficient of variation. It is defined as the ratio of the standard deviation to the mean of a variable $x$, and its formula is $\sqrt{\sum(x_{i}-x_{mean})^2/n}/x_{mean}$. While high scores indicate that members contribute different levels of the specific attribute to the team, low scores indicate that all members contribute similar levels of the attribute to the team. 

The recommendation system normalized all these metrics and summed them into a single metric, the heterogeneity score. While the system calculated this score for each recommendation, including the ones in the Self-Assembled teams' condition, it was only integrated into the Nudged condition's algorithm. 

For our post-hoc statistical analysis, we summed the normalized heterogeneity scores for gender, age, race, internationality, and ethnicity to calculate a ``Surface-Level Heterogeneity Score.'' The ``Deep-Level Heterogeneity Score'' summed the normalized heterogeneity scores for the six project skills. Appendix Table \ref{tab:composition} details the attributes and values that we used to calculate teams' heterogeneity scores. 

To test H2a-b and H3a-b, we conducted two-way ANOVA with agency (present vs. absent) and heterogeneity criteria (included vs. not included) as factors, examining main effects and their interaction for both surface-level and deep-level differences. We used the \texttt{emmeans} package in \textit{R} to estimate marginal means and contrast differences, and we applied a Bonferroni correction for multiple comparisons. We standardized the independent variables prior to analysis and verified normal distribution assumptions. 

\subsubsection{Team Dynamics} 
\paragraph{Perceived Differences.} To assess how salient differences were to team members, participants evaluated how different their teammates were from themselves using Harrison et al.'s \cite{Harrison2002-xd} scales. We computed separate scores for perceived surface-level differences (gender, age, race, ethnicity) and perceived deep-level differences (skills, values, backgrounds). To test H4a-b, we compared perceived surface-level and deep-level differences using two-way ANOVAs that included user agency and heterogeneity criteria.

\paragraph{Communication Frequency} Participants answered sociometric questions identifying the teammates with whom they communicated most frequently during the task. Based on these responses, we computed each participant’s communication indegree (i.e., the number of times a participant was nominated by teammates as a frequent communication partner). To test H5, we compared communication indegree across experimental conditions using a two-way ANOVA that included user agency and heterogeneity criteria. To complement these self-reported measures and to assess whether they aligned with observed interaction patterns, we conducted a transcript-based analysis of teams' conversations. We transcribed Zoom recordings using \textit{Otter.AI}. Trained research assistants reviewed the transcripts to correct transcription errors and assign speaker identities. All transcripts were labeled with participant identifiers, timestamps, and utterance boundaries, and compiled into team-level text files for analysis. Research assistants also conducted a review of the open-ended conversational content to ensure the accuracy and interpretability of the transcripts, as well as to identify salient interaction patterns relevant to communication dynamics. This process was descriptive rather than interpretive.

\subsubsection{Team Outcome} 
Lastly, we recruited ten undergraduate students majoring in marketing and communication to evaluate the teams' projects. We adapted Lykourentzou et al. \cite{Lykourentzou2017-ui}'s evaluation scale and graders rated each team's submission on \textit{originality} and \textit{uniqueness} using 5-point Likert scales (1 = ``Not at all'' to 5 = ``Very much''). Graders worked independently and were blind to experimental conditions. The scale showed high internal consistency (Cronbach's $\alpha = 0.96$), and we averaged the two dimensions to compute each team's creativity score. To test H6, we conducted a two-way ANOVA with agency and heterogeneity criteria as factors, examining the interaction effect. 

%% file: 04_Results.tex
\section{Results}
\label{sec:results}
In total, 332 participants assembled 4-member teams, and 42 individuals worked as solos. Eighty-three teams were assembled: 20 teams in the Random condition, 19 teams in the Assigned-Diverse condition, 25 teams in the Self-Assembled condition, and 19 teams in the Nudged condition. We performed Chi-squared tests of independence (based on 2,000 replicates) to assess whether the proportions of participants' demographic characteristics were equal among conditions. Only the proportion of African American participants statistically differed across conditions ($\chi^2(3, 386)=8.39, p<0.05$).

\subsection{Collaborator Selections}
We begin by analyzing the agency conditions in which participants could choose their collaborators. This analysis operates at the individual level, with each algorithm-generated recommendation as the unit of observation (N = 4,351 recommendations from 185 participants). The mixed-effects model (Table \ref{tab:recommendations_mixed_effects}) confirmed that incorporating heterogeneity criteria into the recommender system's algorithm significantly influenced participants' collaborator choices. As expected, the order of the recommendations (i.e., fit score) strongly predicted participants' selected recommendations ($\beta=2.78, p<0.001$). Therefore, H1a is supported. 

In both conditions, we found that participants were less likely than chance to select collaborators of the opposite gender ($\beta=-0.39, p<0.05$). Including the heterogeneity score had a significant effect on the searches, as participants in the Nudged condition invited fewer individuals than those in the Self-Assembled condition ($\beta=-0.54, p<0.05$). Nevertheless, we found that participants in the Nudged condition were significantly more likely to choose collaborators of a different race ($\beta=0.69, p<0.01$) and different gender ($\beta=0.68, p<0.01$). Thus, H1b and H1c are supported.

\input{recommendations_mixed_effect}

Table \ref{tab:number-recommendations} shows the effect of incorporating heterogeneity criteria into the recommendation algorithm. When heterogeneity criteria were applied, the representation of demographically different collaborators among the top-\textit{N} recommendations increased substantially. For race, the average number of different-race individuals in the top-5 recommendations rose from 2.47 in the Self-Assembled condition to 2.85 in the Nudged condition, and the percentage of invited different-race collaborators increased from 56.03\% to 73.47\%. Similarly, for gender, the average number of different-gender individuals in the top-5 recommendations increased from 1.85 to 2.53, with invitations rising from 40.52\% to 64.29\%. These effects were consistent across values of \textit{N}, indicating that reordering recommendations based on heterogeneity criteria increases the visibility and selection of demographically different collaborators.

\input{number_recommedations}

\begin{figure}[!htb]
\centering
\begin{subfigure}{.47\textwidth}
  \centering
  \includegraphics[width=\linewidth]{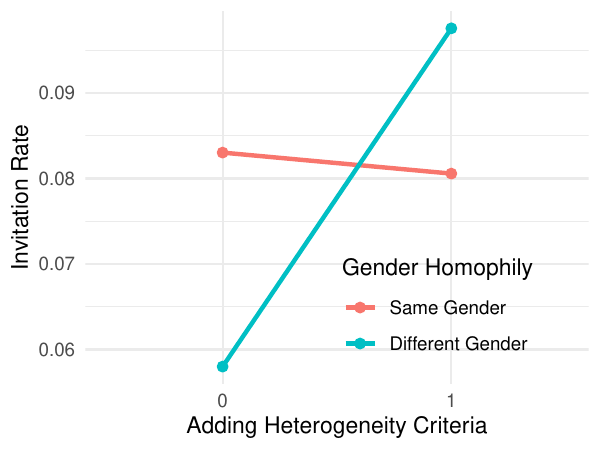}
  \caption{Invitation Rate controlled by Gender Homophily}
  \label{fig:recommendation-effect-gender}
\end{subfigure}%
\hfill
\begin{subfigure}{.47\textwidth}
  \centering
  \includegraphics[width=\linewidth]{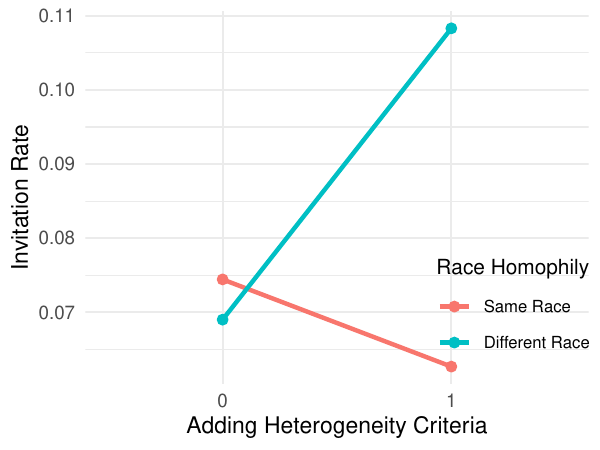}
  \caption{Invitation Rate controlled by Race Homophily}
  \label{fig:recommendation-effect-race}
\end{subfigure}
\caption{Interaction plot between the inclusion of heterogeneity criteria in the algorithm and (a) the gender and (b) race of invited collaborators. The x-axis indicates whether the recommendation algorithm included heterogeneity criteria. The red lines represent average invitation rates for same-gender (same-race), and the blue lines for different-gender (different-race) collaborators.}
\label{fig:recommendation-effect}
\end{figure}

Lastly, Figure \ref{fig:recommendation-effect} illustrates how adding heterogeneity criteria shaped participants' collaborator invitations based on race and gender homophily. While invitation rates for same-race and same-gender collaborators remained stable across conditions, invitations to different-race collaborators increased from 6.9\% to 10.8\%, and invitations to different-gender collaborators increased from 5.80\% to 9.75\% in the Nudged condition, consistent with the regression results showing that incorporating heterogeneity criteria into the algorithm reduced homophily.

\subsection{Team Composition} 
We tested whether user agency reduced team heterogeneity (H2a-b) and whether incorporating heterogeneity criteria increased team heterogeneity (H3a-b), examining both surface-level and deep-level differences. This analysis operates at the team level ($N = 83$ teams), with heterogeneity scores computed from the aggregated team members' attributes.

\begin{figure*}[!htb]
\centering
\begin{subfigure}[b]{.47\textwidth}
\includegraphics[width=\textwidth]{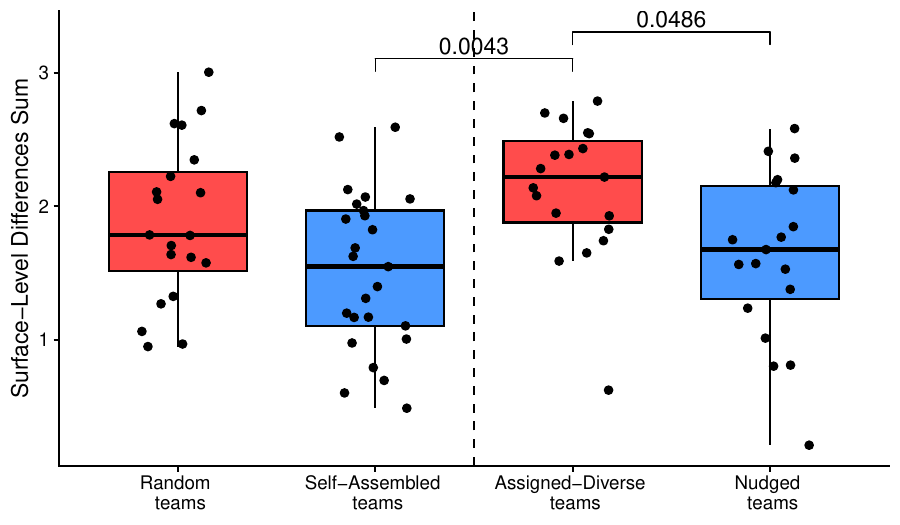}
\caption{Surface-level Heterogeneity Score}
\label{fig:team-composition-surface}
\vspace{2ex}
\includegraphics[width=\textwidth,right]
{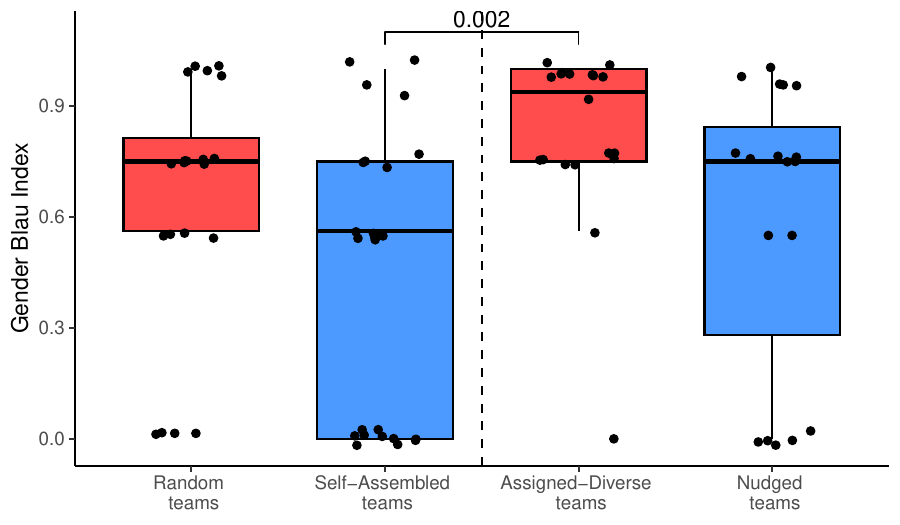}
\caption{Gender Heterogeneity Score (Blau's index)}
\label{fig:team-composition-gender}
\end{subfigure}\qquad
\begin{subfigure}[b]{.47\textwidth}
\includegraphics[width=\textwidth]
{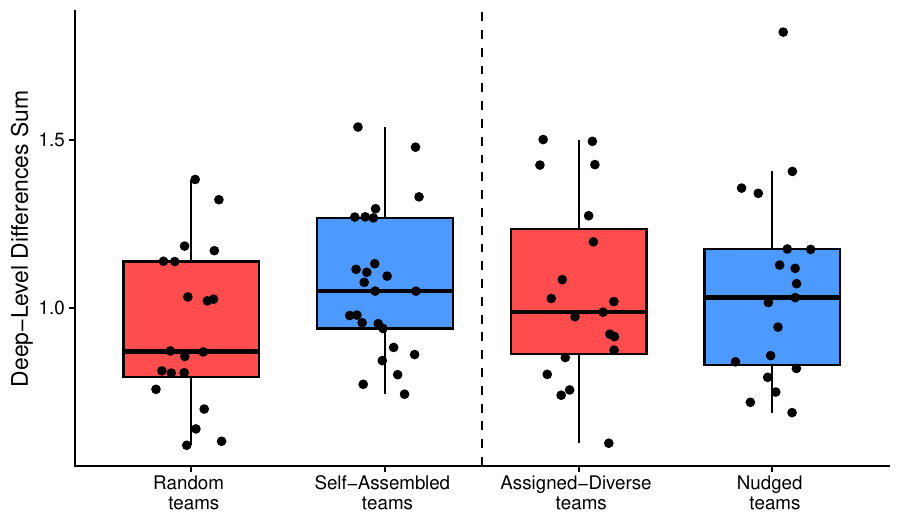}
\caption{Deep-level Heterogeneity Score}
\label{fig:team-composition-deep}
\vspace{2ex}
\includegraphics[width=\textwidth,right]
{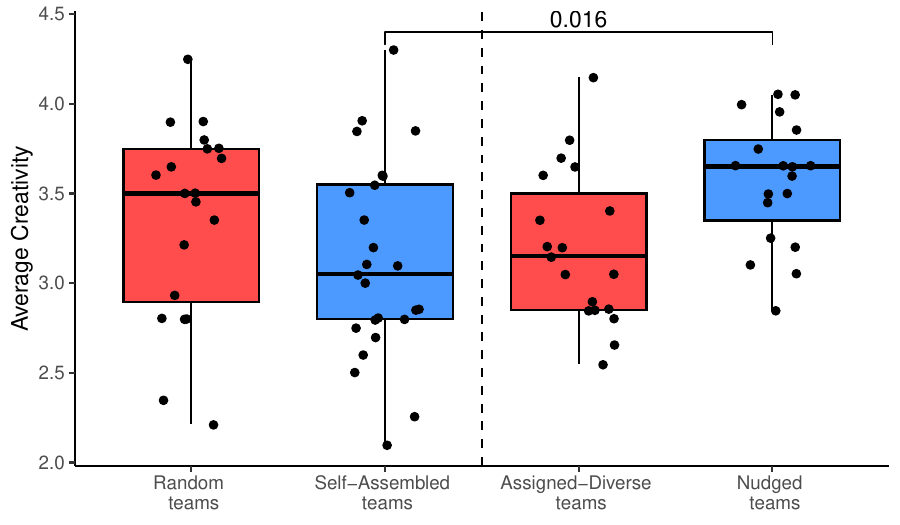}
\caption{Average Creativity Score}
\label{fig:performance}
\end{subfigure}
\caption{\textbf{Results per experimental condition}: (a) Surface-level heterogeneity score is the sum of gender, age, race, and ethnicity heterogeneity scores, (b) Gender heterogeneity was a factor that differed across conditions, (c) Deep-level heterogeneity score is the sum of the six project skills' heterogeneity scores, (d) Average Creativity score measures the recruitment materials' originality and uniqueness. Brackets represent statistically significant differences between two conditions using Tukey HSD tests ($p_{adj} < 0.05$). Each bracket shows its respective \textit{p}-adjusted value. The left panel shows no heterogeneity criteria conditions, while the right panel shows heterogeneity criteria conditions. Red bars represent the no-agency algorithms, while blue bars represent the agency algorithms. Number of observations: 83 teams.}
\label{fig:team-composition}
\end{figure*}

After summing the normalized surface-level difference metrics into a single score, a two-way ANOVA revealed that algorithms incorporating user agency led participants to choose collaborators more similar to themselves, significantly reducing surface-level differences ($F(1,79)=11.53, p<0.01$, Cohen's $\eta^2=0.13$). As shown in Figure \ref{fig:team-composition-surface}, teams in the agency conditions (Self-Assembled: $M=1.51, SD=0.59$; Nudged: $M=1.63, SD=0.62$) exhibited lower surface-level differences than teams in the no-agency conditions (Random: $M=1.87, SD=0.60$; Assigned-Diverse: $M=2.13, SD=0.51$). Post-hoc comparisons confirmed that Assigned-Diverse teams were significantly more heterogeneous than both Self-Assembled teams (Tukey HSD, $\Delta=-0.62, p_{adj}<0.01$) and Nudged teams ($\Delta=-0.50, p_{adj}<0.05$). Therefore, H2a is supported. 

When we analyzed the surface-level difference metrics separately, we found that gender heterogeneity showed the greatest disparity across all conditions (Figure \ref{fig:team-composition-gender}). While agency negatively affected gender balance ($F(1,79)=8.53,p<0.01,\eta^2=0.10$), including heterogeneity criteria in the algorithms significantly helped strengthen it ($F(1,79)=5.39,p<0.05,\eta^2=0.06$). Moreover, the Assigned-Diverse teams achieved the highest gender heterogeneity scores ($M=0.83, SD=0.24$), whereas the Self-Assembled condition ($M=0.43, SD=0.38$) had the lowest scores ($\Delta=-0.40, p_{adj}< 0.01$). As participants could see others' names on the recommendation system, gender became the most salient attribute in their choices. 

In contrast, user agency did not significantly affect deep-level differences. The two-way ANOVA revealed no significant main effect of agency on deep-level differences ($F(1,79)=1.84, p>0.10$). As shown in Figure \ref{fig:team-composition-deep}, there were no significant differences in skill distribution among conditions. A possible explanation is that participants actively sought skill-complementary collaborators regardless of condition. We confirmed this assumption by examining participants' searches on the recommendation system, where the most frequently requested search criteria were project skills, accounting for 49\% of their search queries. Therefore, H2b is not supported.

The two-way ANOVA revealed that incorporating heterogeneity criteria did not significantly increase overall surface-level differences ($F(1,79)=2.08, p>0.10$) or deep-level differences ($F(1,79)=0.62, p>0.10$). Two factors help explain this null result. First, Random teams already achieved moderate heterogeneity levels through chance assignment alone. Indeed, neither surface-level nor deep-level differences between Random and Assigned-Diverse teams were statistically significant. Second, while Nudged teams achieved higher heterogeneity scores than Self-Assembled teams, they scored lower than Random teams, and these differences were not statistically significant. These results suggest that heterogeneity criteria embedded in recommendation algorithms were insufficient to overcome homophily when participants retained agency. Thus, H3a and H3b are not supported.

\subsection{Team Processes}

\paragraph{Perceived Differences Post-Task} 
From the final surveys ($N=332$), we examined whether participants were aware of their surface-level and deep-level differences after the task concluded. A two-way ANOVA reveals that enabling user agency decreased their perceived surface-level differences ($F(1,325)=8.97, p < 0.01$, Cohen's $\eta^2=0.03$). The most significant difference in perceptions was between Assigned-Diverse teams ($M=3.26,SD=0.89$) and Self-Assembled teams ($M=3.0, SD=0.77$), with a difference of 8\% ($p_{adj}<0.05$). Thus, H4a is supported. We found that user agency had no significant effect on participants' perceived deep-level differences (Two-way ANOVA, $F(1,325)=0.73, p>0.10$). Overall, participants reported being similar in terms of skills and values scores, even after interacting for approximately 30 minutes. This null result may be explained by the similarity in backgrounds and profiles within teams, as we did not recruit participants based on their disciplines or areas of knowledge. Thus, H4b is not supported.

\paragraph{Communication Metrics}
We found that the team members' communication network indegree was significantly affected by the inclusion of heterogeneity criteria (Two-way ANOVA, $F(1,325)=9.18$, Cohen's $\eta^2=0.03$). Specifically, participants in the Assigned-Diverse teams reported communicating with fewer teammates compared to the Random teams' participants ($\Delta=-0.52, p_{adj} < 0.01$) and Self-Assembled teams ($\Delta=-0.47, p_{adj} < 0.05$). This result indicates that the highly heterogeneous teams assigned by the algorithm struggled with communication, as some members did not participate in the conversation, may not have known how to initiate their conversations, or did not work together effectively.  Notably, while Nudged teams also incorporated heterogeneity criteria, their communication levels did not differ significantly from Random or Self-Assembled teams, consistent with the marginal interaction effect. This pattern suggests that algorithmically assigning members to heterogeneous teams created communication barriers, as some members did not participate actively or struggled to initiate collaboration with unfamiliar teammates. In contrast, allowing users to choose their collaborators, even when nudged toward heterogeneity, mitigated these challenges. Thus, H5 is supported.

We confirmed these results by analyzing participants' open-ended responses. In the Assigned-Diverse teams, we found several critiques of their communication and alignment, such as ``\textit{coordination, getting on the same page};'' ``\textit{Very scattered. Not everyone was on the same page. Started to quickly without a defined goal.};'' or ``\textit{Initial kick-off on who was going to be the lead and what the approach and presentation needed to communicate.}'' Another participant reported that ``\textit{I really wanted to share my ideas but I didn't feel comfortable and I felt like they would not have listened to me}.'' Although participants in the Self-Assembled teams reported higher levels of comfort, some relevant critiques were ``\textit{[having a] single vision};'' ``\textit{Making our ideas look pretty};'' and ``\textit{All the rest were younger and have similar styles (but none had experience working with design).}'' In the case of the Nudged teams, relevant critiques were the lack of designers in the team and establishing consensus (e.g., ``\textit{...[having] different backgrounds sometimes made it hard for us to figure out what we wanted to prioritize, which led to each of us designing our own marketing materials}'').

\subsection{Team Outcomes}
In our last analysis of the team deliverables ($N=83$), we find that the team formation algorithms led to significant differences in team outcomes (Figure \ref{fig:performance}). The two-way ANOVA revealed that neither agency ($F(1,79)=0.08, p>0.10$) nor heterogeneity criteria ($F(1,79)=2.21, p>0.10$) alone significantly affected team creativity. However, the interaction between agency and heterogeneity criteria was significant ($F(1,79)=8.23, p<0.01, \eta^2=0.09$), indicating that the effect of heterogeneity criteria depended on whether users had agency over collaborator selection.

Tukey HSD tests revealed that Nudged teams achieved significantly higher scores than Self-Assembled teams ($\Delta=0.45, p_{adj}<0.05$, Cohen's $\eta^2=0.09$). No other pairwise comparisons reached statistical significance, though Nudged teams also scored higher than Assigned-Diverse teams at a marginal level ($\Delta=0.37, p_{adj}>0.10$). Notably, neither agency nor heterogeneity criteria alone improved outcomes relative to Random teams: Self-Assembled teams scored slightly lower ($\Delta=-0.24, p_{adj}>0.10$) and Assigned-Diverse teams showed no meaningful difference ($\Delta=-0.16, p_{adj}>0.10$). Thus, H6 is supported.

These results suggest that team outcomes depended on how agency and heterogeneity criteria were combined. When participants chose collaborators without receiving team composition guidance (i.e., Self-Assembled), they formed more homogeneous teams that underperformed for this task. When heterogeneity was optimized without user input (i.e., Assigned-Diverse), teams became more heterogeneous but faced communication barriers that limited their effectiveness. Nudging offered a different balance: it embedded heterogeneity criteria within a choice architecture that preserved user agency, allowing teams to benefit from a more varied composition while avoiding some of the communication costs observed in assigned teams. 

%% file: recommendations_mixed_effect.tex
\begin{table}[!htb]
\centering
\resizebox{0.7\columnwidth}{!}{%
\begin{tabular}{lccc}
\toprule
 & \textbf{Estimate} & \textbf{S.E.} & \textbf{z-value}    \\ \midrule
\textit{Fixed effects}  &          &  \\
Intercept  & -5.31***    & 0.39 & -13.65\\
H1a: Rank Order (i.e., Fit Score)  & 2.78*** & 0.32 & 8.80 \\
Different race & -0.01 & 0.17 & -0.04 \\
Different gender   & -0.39*  & 0.17 & -2.32  \\
Including heterogeneity criteria (i.e., Treatment) & -0.54*    & 0.27 & -2.00   \\
H1b: Different race $\times$ Treatment    & 0.69** & 0.26 & 2.63 \\
H1c: Different gender $\times$ Treatment    & 0.68** & 0.25 & 2.66 \\ \bottomrule
\textit{Random effects} & \textbf{Variance} & \textbf{S.D.}    \\ \midrule
Sender & 0.64     & 0.80    \\ \bottomrule
\end{tabular}%
}
\caption{Generalized linear mixed-effects logistic regression predicting participants’ invitations across agency conditions. The interaction term shows that when heterogeneity criteria were included in the algorithm, participants were more likely to invite collaborators of different genders and races. Number of observations: 4,351 recommendations. Groups (Random variable): Participants that searched for collaborators (N=185); BIC=2,249.4; logLik: -1,116.7. Variables were standardized before running the model. Stars indicate the statistical significance of each coefficient: * \textit{p} < 0.05, ** \textit{p} < 0.01, *** \textit{p} < 0.001.}
\label{tab:recommendations_mixed_effects}
\end{table}

%% file: number_recommedations.tex
\begin{table}[!htb]
\resizebox{\columnwidth}{!}{%
\begin{tabular}{
    >{\centering\arraybackslash}p{0.10\textwidth}|
    >{\centering\arraybackslash}p{0.09\textwidth}
    >{\centering\arraybackslash}p{0.09\textwidth}|
    >{\centering\arraybackslash}p{0.09\textwidth}
    >{\centering\arraybackslash}p{0.09\textwidth}
    >{\centering\arraybackslash}p{0.09\textwidth}|
    >{\centering\arraybackslash}p{0.09\textwidth}
    >{\centering\arraybackslash}p{0.09\textwidth}|
    >{\centering\arraybackslash}p{0.09\textwidth}
    >{\centering\arraybackslash}p{0.09\textwidth}
    >{\centering\arraybackslash}p{0.09\textwidth}
}
\toprule
\multicolumn{1}{c}{\textbf{\shortstack{Number of \\ top-N \\ recs.}}} & \multicolumn{2}{c}{\textbf{\shortstack{Average number of \\ different-race \\ displayed recs.}}} & \multicolumn{3}{c}{\textbf{\shortstack{Percentage of \\ different-race\\ invited recs.}}} & \multicolumn{2}{c}{\textbf{\shortstack{Average number of \\ different-gender \\ displayed recs.}}} & \multicolumn{3}{c}{\textbf{\shortstack{Percentage of \\ different-gender\\ invited recs.}}} \\ \midrule
 & Self-A. & Nudged & Self-A. & Nudged & Diff. & Self-A. & Nudged & Self-A. & Nudged & Diff. \\ \hline
3 & 1.29 & 1.58 & 37.07\% & 51.02\% & +13.95\% & 0.95 & 1.31 & 27.59\% & 41.84\% & +14.25\%\\
4 & 1.84 & 2.14 & 45.69\% & 64.29\% & +18.60\% & 1.39 & 1.90 & 32.76\% & 56.12\% & +23.36\%\\
5 & 2.47 & 2.85 & 56.03\% & 73.47\% & +17.43\% & 1.85 & 2.53 & 40.52\% & 64.29\% & +23.77\%\\
6 & 3.16 & 3.48 & 65.52\% & 80.61\% & +15.10\% & 2.28 & 3.03 & 43.97\% & 68.37\% & +24.40\%\\
7 & 3.81 & 4.11 & 72.41\% & 85.71\% & +13.30\% & 2.76 & 3.54 & 48.28\% & 72.45\% & +24.17\%\\
8 & 4.47 & 4.67 & 76.72\% & 88.78\% & +12.05\% & 3.20 & 3.99 & 50.86\% & 73.47\% & +22.61\%\\
9 & 5.13 & 5.29 & 81.90\% & 92.86\% & +10.96\% & 3.66 & 4.43 & 53.45\% & 74.49\% & +21.04\%\\
10 & 5.74 & 5.89 & 83.62\% & 93.88\% & +10.26\% & 4.11 & 4.81 & 55.17\% & 75.51\% & +20.34\%\\ \bottomrule
\end{tabular}%
}
\caption{Race and gender representation in the top-N recommended collaborators. The table shows the average number and percentage of different-race and different-gender individuals appearing in the top-N recommendations (recs.) for the Self-Assembled (control) and Nudged (treatment) conditions.}
\label{tab:number-recommendations}
\end{table}

%% file: 05_Discussion.tex
\section{Discussion}
\label{sec:discussion}
This study empirically examined how algorithmic interventions shape team formation, from initial collaborator selections through team processes to performance outcomes (Table \ref{tab:summary}). Through a laboratory experiment involving over 320 participants, we tested the effects of user agency and heterogeneity criteria---separately and in combination---on teams assembled via a recommendation system. Our findings reveal that the mechanism by which heterogeneity is achieved matters as much as the outcome.

\input{summary}

\subsection{Nudging Collaborator Choices}
This study extends previous research on team formation algorithms by showing that re-ranking recommendations with heterogeneity criteria can meaningfully shift collaborator choices (H1a-c). Specifically, integrating a heterogeneity score into the recommendation algorithm significantly improved gender and race representation among the top-ranked suggestions. One likely reason is that demographic homophily---a primary tendency in team formation \cite{Hinds2000-uu,Lungeanu2014-va}---emerges naturally in recommendation settings, as users infer race and gender from names or profile pictures \cite{Jahanbakhsh2020}. In contrast, we did not observe significant effects on deep-level differences. A plausible explanation is the relative homogeneity of the participant pool: participants shared broadly similar educational backgrounds, and recruitment was not stratified by profession or disciplinary expertise. As a result, meaningful differences in cognitive perspectives or domain knowledge may have been limited. Despite this constraint, rebalancing recommendations to account for heterogeneity increased the visibility of demographically different collaborators at the top of the list, shaping team composition.

While previous research in CSCW has demonstrated that explicit composition-related information---such as scores or visual badges \cite{Gomez-Zara2020-ol}---can paradoxically exacerbate users' attachment to similar individuals, our findings reveal that simply reordering algorithmic recommendations can subtly counteract homophily without making those criteria salient to users. This distinction is theoretically important because different mechanisms of influence can shape collaboration opportunities in different ways. Highlighting demographic and cognitive differences can potentially trigger social categorization processes that reinforce in-group preferences \cite{Tajfel1979-yu,TWYMAN202284}. In contrast, nudging through re-ranking operates through the choice architecture itself, changing which potential collaborators receive attention without revealing the criteria behind the ranking.

At the same time, this opacity introduces several risks. Because nudging operates without users' explicit awareness, its effect depends critically on how the nudge is designed and deployed. In our study, nudging functioned as an intermediate approach between free choice and assignment, producing outcomes that differed from both extremes. However, poorly calibrated, misaligned, or harmful nudges could just as easily amplify existing preferences or channel attention in ways that reinforce rather than mitigate group segregation \cite{bovens2009ethics,Chaney2018,yeung2019hypernudge}. These findings underscore that algorithmic nudging has consequences that hinge on design choices that govern what---and who---becomes visible within users' attention frames.

\subsection{The Composition-Process Trade-off}
Our findings suggest a trade-off between optimizing team composition for heterogeneity and enabling that heterogeneity to translate into effective team functioning, extending prior work on the ``double-edged sword'' of team heterogeneity \cite{Horwitz2007-fu,Bell2011-wu}. Rather than focusing solely on whether varied team composition affects outcomes, our study shows that the mechanism through which that composition is shaped also matters.

For instance, the Assigned-Diverse teams achieved the highest heterogeneity levels but experienced corresponding process costs. Their team members reported the highest awareness of demographic differences and the lowest communication frequency across the conditions. This pattern suggests that algorithmic assignment may trigger the negative dynamics associated with heterogeneous teams, such as intergroup anxiety, social categorization, and coordination difficulties \cite{Harrison1998-hs,Lau2005-pb}. Prior work similarly shows that individuals resist externally assigned placements and become more aware of demographic differences when composition is imposed rather than chosen \cite{Dobbin2016-xt,Dover2016-xx}. These effects may depend less on actual composition alone than on how members perceive and experience it \cite{Shemla2016-vs}, suggesting that perceived autonomy in team formation may be as important as actual composition for subsequent collaboration quality.

In contrast, nudging appeared to create a more productive balance. Nudged teams delivered the most creative projects, even though their heterogeneity levels were not significantly higher than those of Self-Assembled teams. This suggests that nudging may have introduced enough variation to enable heterogeneity benefits while preserving the sense of choice needed for team commitment and satisfaction. Nudged teams achieved moderate heterogeneity without elevated awareness of differences or reduced communication. The act of choosing---regardless of whether choices were nudged---appears to create psychological ownership that buffers against intergroup tensions \cite{Chapman2006}. 

\subsection{Calibrating User Agency}
Our findings highlight that both granting and restricting user agency introduce distinct costs. When users retain agency, collaboration choices tend to reflect individual preferences, with homophily emerging as a dominant organizing principle. In socio-technical systems, this tendency can be amplified by algorithmic mediation, as recommendation systems often reinforce similarity and like-minded communities \cite{pariser2011filter}. These dynamics suggest that agency, while normatively desirable, can undermine collective goals when left entirely unconstrained.

Conversely, restricting agency through algorithmic assignment shifts coordination costs in a different direction. When team composition is externally imposed, demographic and cognitive differences are likely to become more salient, potentially heightening social categorization and inhibiting interaction. Allowing individuals to participate in collaborator selection---even within a structured or nudged environment---appears to mitigate these effects by fostering a sense of psychological ownership and control \cite{Shemla2016-vs}. Taken together, these findings suggest that agency is not simply a design choice to maximize or minimize. Instead, effective team formation systems need to calibrate agency as a design lever that balances individual autonomy with collective organizational values and goals \cite{Heer2019,Bennett2023-qw}.

\subsection{Ethical Implications}
The capacity of algorithms to shape team composition raises questions extending beyond technical design to core organizational values. Our findings demonstrate that algorithmic design choices have tangible consequences for team composition, communication, performance, and the overall experience of individuals in their collaborations.

\subsubsection{Delegating Organizational Values to Algorithms}
As exemplified in this study, algorithmic team formation entails delegating organizationally value-laden decisions to computational systems. When algorithms optimize team composition, they embed contestable assumptions about which attributes should matter, how those attributes should be measured, and whether algorithmic intervention is an appropriate means of pursuing organizational goals \cite{Fazelpour2022-il}. These assumptions are not neutral technical choices; they reflect organizational values that may diverge from individual preferences or alternative conceptions of fairness.

Delegating team formation to algorithms, therefore, does not eliminate value judgments but redistributes them, from individuals and managers to system designers and organizations \cite{Polzer2022-pl}. Our findings show that this redistribution has material consequences as different algorithmic designs (i.e., assignment, nudging, or free choice) enact organizational values in distinct ways, shaping not only who collaborates with whom, but also how teams experience coordination and communication. Importantly, these effects emerged even when users were not explicitly aware that heterogeneity criteria were influencing their options.

This delegation also raises concerns of legitimacy and contestability. In nudging-based systems, influence operates by shaping attention and salience rather than by explicit constraint, blurring the boundary between individual choice and organizational control. Ethical algorithmic team formation, therefore, requires not only careful technical design and testing, but also governance mechanisms that support transparency, explanation, and avenues for contestation, especially when such systems shape access to collaboration and opportunity over time \cite{alfrink2023contestable}. Our results suggest that ethical evaluation of algorithmic team formation must attend not only to outcomes, but to the mechanisms through which organizational values are enacted in everyday collaborative decisions.

\subsubsection{Transparency and Hidden Manipulation}
While nudging heterogeneity through algorithmic re-ranking produced benefits for some individuals, it raises distinct concerns about transparency and autonomy \cite{rader2018explanations,sankaran2020respecting,gray2018dark}. Participants in our Nudged condition experienced their choices as autonomous while their options were being shaped by criteria they did not choose and could not see. This hiddenness may be essential to nudging's effectiveness, as it creates a perception of autonomy that likely contributes to psychological ownership. Yet when invisible nudges are embedded in socio-technical systems, individuals cannot perceive or contest them \cite{caraban201923,Raveendhran2021-lc}. Previous research suggests transparency about algorithmic influence can undermine effectiveness while increasing perceived legitimacy \cite{logg2019algorithm}. This creates difficult design questions: Should users be told recommendations are adjusted to form teams with varied skills and backgrounds? Or would disclosure trigger the backlash effects observed when people are aware of their differences \cite{Gomez-Zara2020-ol}? 

Regardless of the designers' benevolent purposes, opaque algorithmic architectures further limit users' trust, interpretability, and the ability to refine or contest recommendations. Such interventions can systematically favor some individuals while disadvantaging others over time, reinforcing inequality and biases \cite{noble2018algorithms}. While our system did not employ any particular machine learning model to infer people's preferences, the algorithmic nature of decision-making can still make explanations and information more difficult to provide. These challenges were particularly evident in the decision not to display participants' information about their team composition, as many individuals have opposite reactions toward unfamiliar and different people \cite{Gomez-Zara2020-ol,Dobbin2016-xt}. Providing context-sensitive explanations, along with options to opt out of recommendation adjustments, may increase receptivity and legitimacy of these interventions \cite{gedikli2014should,gray2018dark}.  

\subsubsection{Algorithms as Social Gatekeepers in Collaborative Work}
Our study provides empirical evidence that social recommendation algorithms can shape who people choose to work with, even without explicit cues or disclosures. In doing so, algorithmic nudges become consequential socio-technical artifacts that influence not only task coordination, but also the relational, social, and identity dimensions of collaborative work \cite{Kellogg2020-oo}. Unlike many traditional recommender systems that guide low-stakes or ephemeral choices, team formation systems mediate access to collaboration opportunities that structure future reputation, skill development, and professional trajectories. As a result, their effects extend to who is interviewed, hired, and promoted over time \cite{schumann2020we}.

Viewing algorithmic team formation through this lens foregrounds its role as a mechanism of \textit{social gatekeeping} \cite{granovetter1973strength,burt2002social,bozdag2013bias}. By re-ranking collaborator recommendations, our intervention shifted exposure in the top-ranked results toward a more representative set of potential collaborators, aligning with prior work on fairness-aware ranking and exposure \cite{Geyik2019-qt}. At the same time, this redistribution of visibility raises concerns about fairness in how benefits and burdens are allocated. Some potential collaborators may receive less exposure in the ranking, while others may be repeatedly surfaced based on demographic attributes they did not seek or consent to foreground.

Our process-level findings further complicate these trade-offs. If heterogeneous teams experience greater discomfort or communication barriers, then composition-oriented interventions may impose psychological or interactional costs on some individuals in pursuit of collective or organizational goals. Because teams shape long-term professional outcomes, these costs may compound over time in ways that are difficult to anticipate or reverse \cite{schumann2020we}. Delegating team formation to algorithms thus entails not only compositional decisions, but also ethical responsibility for their downstream and longitudinal consequences.

Finally, our findings caution against delegating judgments about what is ``best'' for teams entirely to algorithmic optimization. Increasing heterogeneity and visibility are valuable, but they do not guarantee effective collaboration. Both Assigned-Diverse and Nudged teams in our study encountered communication and coordination challenges, underscoring the difficulty of aligning compositional goals with high-quality team processes. While preserving user agency mitigated some of these tensions, algorithmic nudging represents only one of many possible design choices. Future systems should move beyond narrow demographic optimization toward more holistic approaches to team composition that consider skills, relational history, and support structures \cite{Gomez-Zara2020-ol}. Moreover, algorithmic interventions may need to be complemented by organizational practices---such as mentoring, leadership development, and team-building---to help teams translate algorithmically shaped composition into productive and sustainable collaboration \cite{Hastings2018,Page2019-dm}.

\subsection{Limitations and Future Work}
It is important to acknowledge the limitations of this study. First, the experiment was conducted in a laboratory setting, using a fictitious scenario and a short creativity task. Although laboratory experiments increase internal validity by isolating collaborative dynamics from confounds present in real-world settings, teams assembled in laboratory experiments do not have a past or a future. The 30-minute task also limits what we can observe about how teams develop over time, especially because heterogeneous teams may require more time and support to function effectively. Future research, including field and quasi-experimental studies, should examine whether these findings generalize to real-world, longer-term teams working on consequential organizational tasks.

Second, participants completed only one type of task. Creativity tasks are common in CSCW and team research \cite{Lykourentzou2017-ui,Dow2011,salehi2018}, but the effects of team formation algorithms may vary across task contexts. In particular, compositional differences may matter differently for creativity, decision-making, problem-solving, or execution-oriented tasks \cite{mcgrath1984groups}. Future studies should examine how different team formation strategies perform across a broader range of collaborative tasks. 

Third, our participant pool and experimental sessions may limit generalizability. We recruited participants from a local area, many of whom were university students and staff. We also conducted several medium-sized sessions rather than a large-scale session for each experimental condition, which allowed participants enough time to review other profiles and recommendations. However, these choices may have introduced session-level variation. Future research should conduct larger-scale sessions with more varied participant pools, including participants recruited from online experiment platforms or organizational settings.

Fourth, the design of the recommendation system shaped the results. How we operationalized and computed heterogeneity affected the recommendations produced by the algorithms \cite{Fazelpour2022-il}. As discussed in \cite{Gomez-Zara2020-ol}, team heterogeneity is more complex than a single score, and future work should test alternative metrics and combinations of attributes to assess their effects on collaborator choices and resulting teams. Another design choice in this experiment was to allow participants to assemble their teams sequentially: two participants would agree to form a team, then invite others to join. Because teams can form in more complex ways, the order and current status of a team could have influenced participants' collaborator decisions. Moreover, participants could only set their collaboration preferences on the system, rather than searching for names or keywords. Future work should examine alternative team formation procedures. 

Fifth, the information available to participants was limited and shaped their decisions. Participants saw their real names on the platform, which may have shaped social inferences, but they did not see survey responses, connections, references, resumes, or other information commonly available in real-world collaboration platforms. As a result, participants may have relied heavily on names, profile pictures, and rankings when selecting collaborators. Future research should test more realistic platforms and investigate how different types of information affect users' responses to team formation recommendations.

Finally, participants were not informed about the algorithmic interventions before the experiment started. The system did not explain how the team formation algorithm worked to avoid influencing participants' decisions. Providing explanations, disclosing the use of heterogeneity criteria, and priming participants with the benefits of different team formation strategies could yield different outcomes. Researchers adopting similar approaches should carefully consider how much participants are informed about algorithmic interventions, how organizational values are embedded into system design, and what safeguards are needed when team formation affects real-world collaboration or career-relevant outcomes.

%% file: summary.tex
\begin{table*}[!htbp]
\centering
\caption{Summary of Hypotheses and Results}
\label{tab:summary}
\resizebox{\columnwidth}{!}{%
\begin{tabular}{p{0.8cm} p{7.5cm} p{2.5cm} p{5.5cm}}
\toprule
\textbf{H} & \textbf{Hypothesis} & \textbf{Result} & \textbf{Key Finding} \\
\midrule
\multicolumn{4}{l}{\textit{\textbf{Collaborator Selection} (Individual-level, N = 4,351 recommendations)}} \\
\addlinespace[0.5ex]
H1a & Higher-ranked recommendations will be more likely to be selected & Supported & Fit score strongly predicted selection. \\
\addlinespace[0.5ex]
H1b & Heterogeneity criteria will increase the selection of different-race collaborators & Supported & Increased different-race selections. \\
\addlinespace[0.5ex]
H1c & Heterogeneity criteria will increase the selection of different-gender collaborators & Supported & Increased different-gender selections. \\
\midrule
\multicolumn{4}{l}{\textit{\textbf{Team Composition} (Team-level, N = 83 teams)}} \\
\addlinespace[0.5ex]
H2a & User agency will reduce surface-level differences & Supported & Agency reduced surface-level differences. \\
\addlinespace[0.5ex]
H2b & User agency will reduce deep-level differences & Not supported & N/A \\
\addlinespace[0.5ex]
H3a & Heterogeneity criteria will increase surface-level differences & Not supported & N/A \\
\addlinespace[0.5ex]
H3b & Heterogeneity criteria will increase deep-level differences & Not supported & N/A \\
\midrule
\multicolumn{4}{l}{\textit{\textbf{Team Processes} (Individual-level, N = 332 participants)}} \\
\addlinespace[0.5ex]
H4a & Assigned-Diverse teams will report higher perceived surface-level differences & Supported & Agency reduced perceived differences. \\
\addlinespace[0.5ex]
H4b & Assigned-Diverse teams will report higher perceived deep-level differences & Not supported & N/A \\
\addlinespace[0.5ex]
H5 & Assigned-Diverse teams will report less frequent communication & Supported & Heterogeneity criteria reduced communication. \\
\midrule
\multicolumn{4}{l}{\textit{\textbf{Team Outcomes} (Team-level, N = 83 teams)}} \\
\addlinespace[0.5ex]
H6 & Combining agency and heterogeneity criteria will produce better outcomes than either alone & Supported & Significant interaction effect; Nudged outperformed Self-Assembled \\
\bottomrule
\end{tabular}}
\end{table*}

%% file: 06_Conclusion.tex
\section{Conclusion}
\label{sec:conclusion}
This study advances our understanding of algorithms in organizational contexts by showing how design choices regarding user agency and organizational values systematically influence team composition, interaction processes, and collaboration outcomes. Our findings demonstrate that algorithmic team formation is not a neutral technical intervention. Different algorithmic designs produce meaningfully different patterns of team composition, acceptance, communication, and performance. In particular, assignment, nudging, and free choice represent distinct governance regimes that redistribute decision-making power between individuals and organizations in consequential ways.

As algorithms increasingly mediate who works with whom, the choices embedded in these systems---what to optimize, how much agency to preserve, and whether and how to intervene---become organizational decisions with ethical and social implications. These decisions shape not only short-term team effectiveness but also access to collaboration opportunities, professional trajectories, and collective capacity over time. We argue that greater attention is needed to understand the role of algorithms as social gatekeepers, how organizational values are encoded into algorithmic systems, how these values are enacted in practice, and how they can be governed responsibly. Understanding these dynamics will be essential for designing algorithmic systems that do more than coordinate work efficiently. Such systems should enable people to discover collaborators, form teams they can trust, and build the relationships through which collective work becomes possible.

%% file: 07_Appendix.tex

\newpage
\appendix
\label{appendix}
\setcounter{table}{0}
\setcounter{figure}{0}
\renewcommand{\thefigure}{\thesection.\arabic{figure}}    
\renewcommand{\thetable}{\thesection.\arabic{table}}

\section{Supplementary Images}

\begin{figure*}[!htb]
    \centering
    \begin{subfigure}[b]{.99\textwidth}
    \includegraphics[width=\textwidth,trim={0 2cm 0 0},clip]{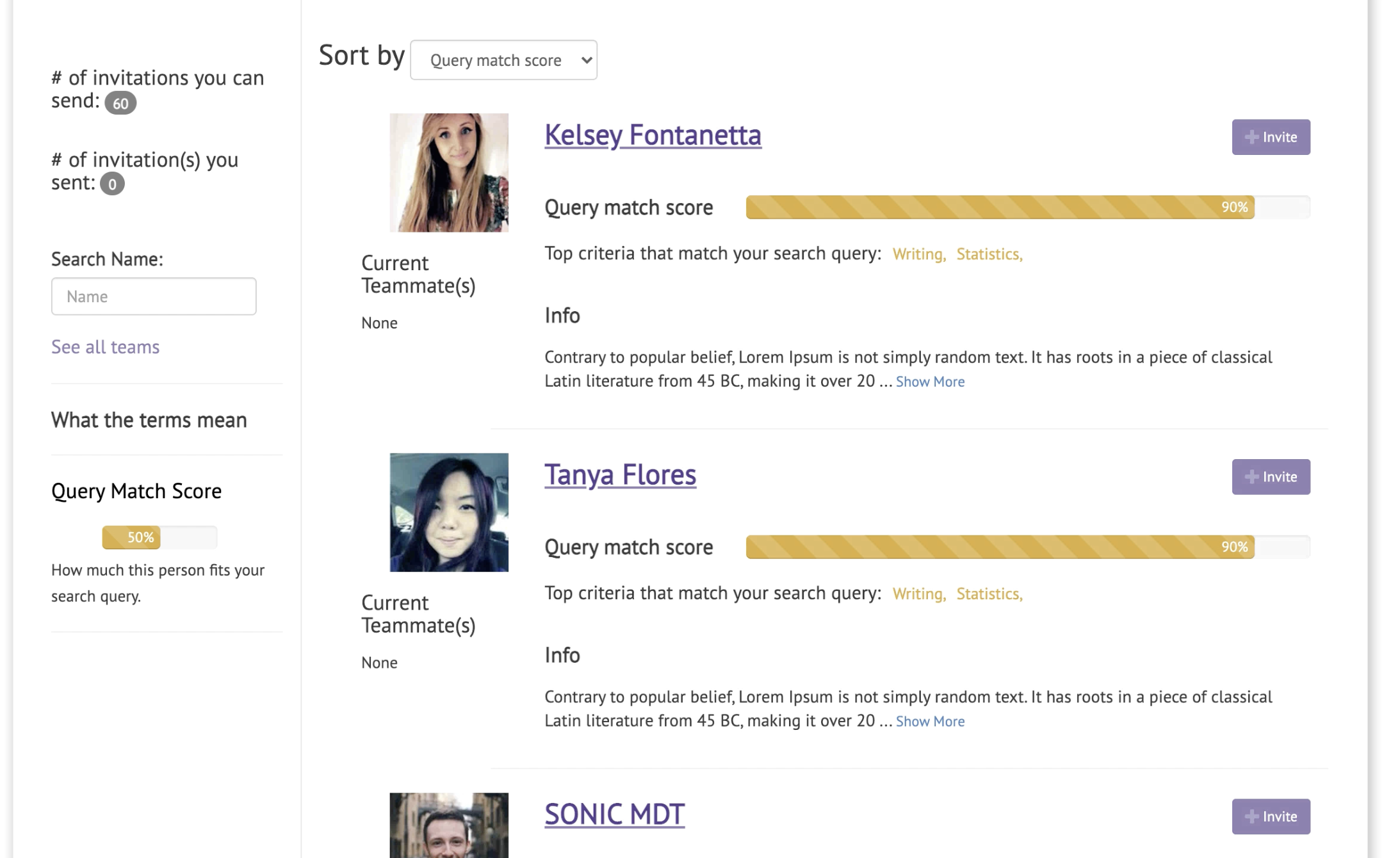}
    \caption{The dashboard displays a list of recommended team members.}
    \label{fig:recommender-system-list}
    \vspace{2ex}
    \includegraphics[width=\textwidth,right]{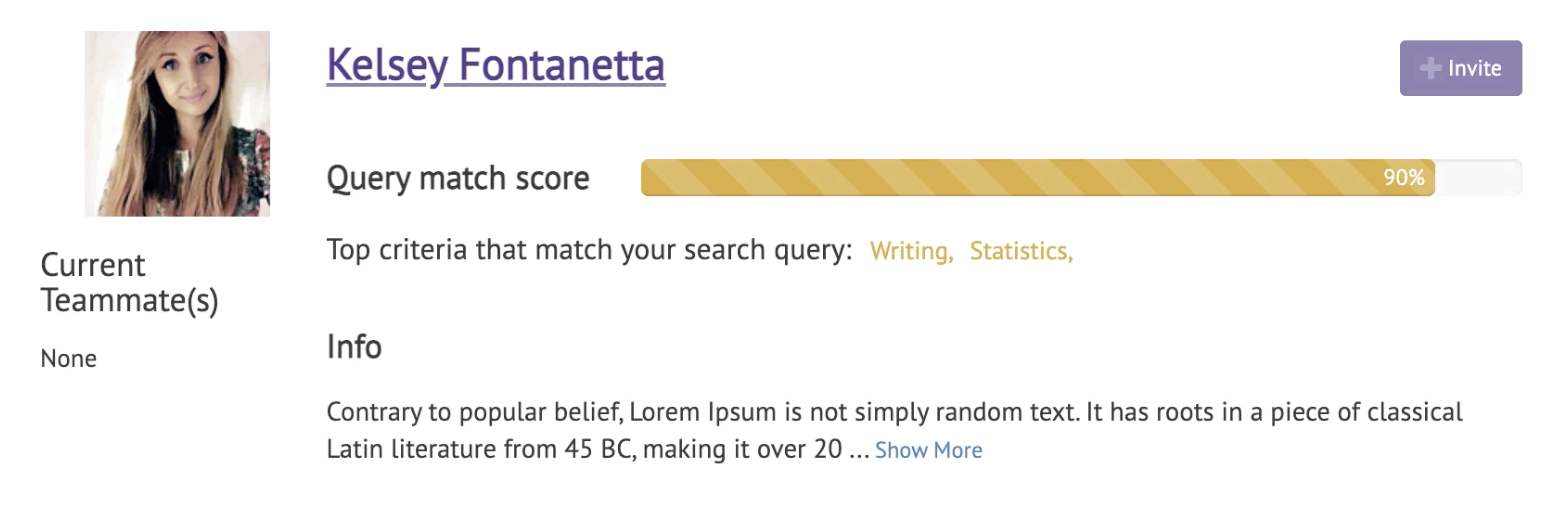}
    \caption{Example of a potential collaborator recommendation. The recommendation displayed a photo of the team member candidate (or a neutral avatar if the picture was not uploaded), the full name of the participant, a query match score indicating the fit of the candidate to the searcher’s query, the top criteria that matched the search criteria, and a brief display of the candidate’s public profile. The recommendation system displayed an invitation button in the top-right corner of each recommendation.}
    \Description{Example of a potential collaborator recommendation. The recommendation displayed a photo of the team member candidate (or a neutral avatar if the picture was not uploaded), the full name of the participant, a query match score indicating the fit of the candidate to the searcher’s query, the top criteria that matched the search criteria, and a brief display of the candidate’s public profile. The recommendation system displayed an invitation button in the top-right corner of each recommendation.}
    \end{subfigure}
    \caption{\textbf{Recommendation examples display the ``My Dream Team'' Recommender. Screenshot from the authors’ system.}}
    \label{fig:recommender-system}
\end{figure*}

\begin{figure}[!htb]
    \centering
    \includegraphics[width=\textwidth]{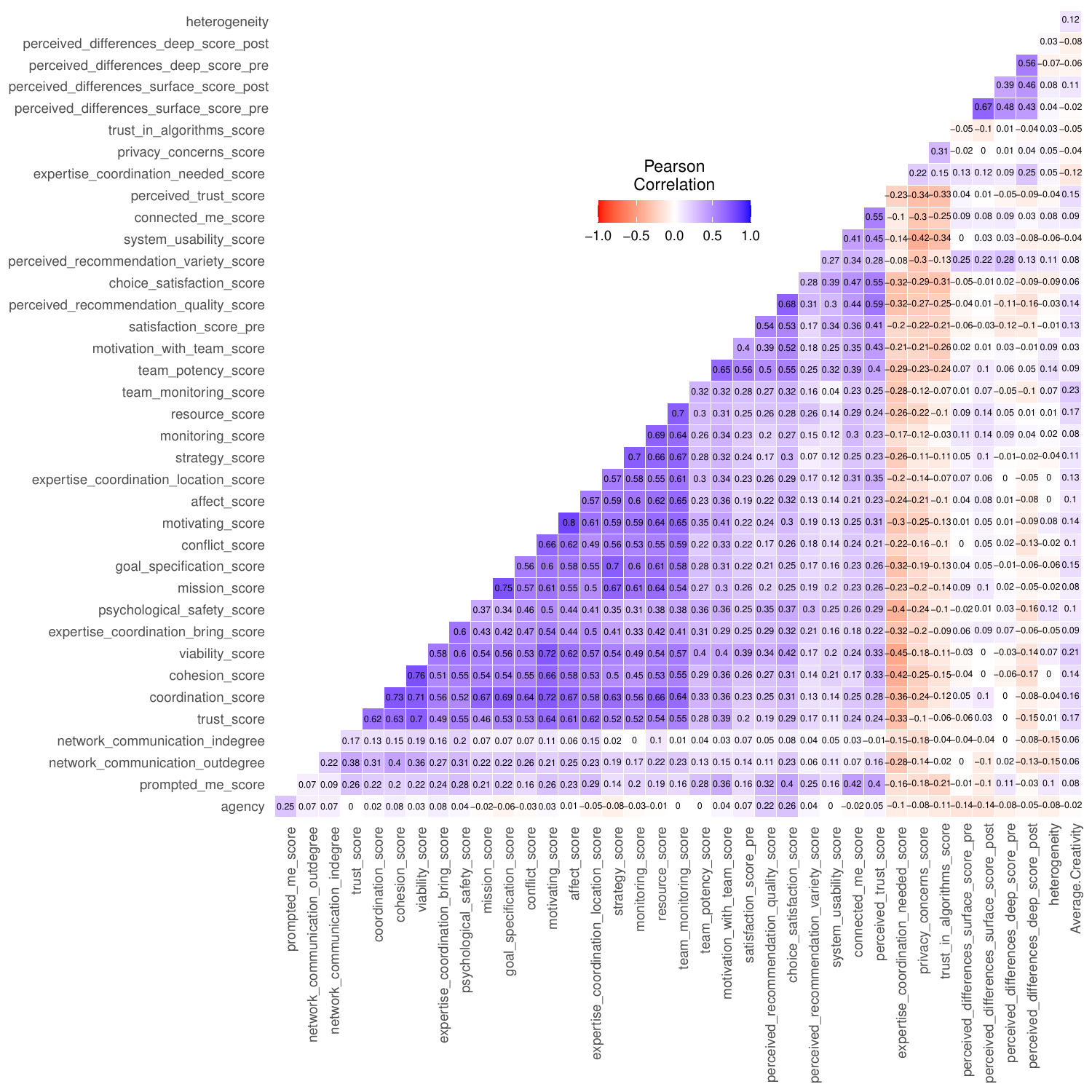}
    \caption{Correlation among variables}
    \Description{A heatmap that illustrates the correlation matrix of various metrics related to team dynamics, system usability, and individual perceptions, with each cell representing the correlation coefficient between two variables. The color gradient ranges from dark purple/blue for strong positive correlations to light shades for weaker correlations, and red for negative correlations. The variables, including aspects like trust, diversity, psychological safety, and team potency, are compared against each other, highlighting the relationships between them. The diagonal shows perfect correlations (1.0), as each variable is compared to itself.}
    \label{fig:appendix:correlations}
\end{figure}

\clearpage

\section{Supplementary Tables}

\input{composition_attributes}
\input{post_assembly_survey}
\input{final_survey}


%% file: composition_attributes.tex
\begin{table}[!htb]
\centering
\small
\renewcommand{\arraystretch}{2}
\resizebox{\textwidth}{!}{%

\begin{tabular}{p{0.25\textwidth}p{0.50\textwidth}p{0.25\textwidth}}
\toprule
\textbf{Metric} &
  \textbf{Scale} &
  \textbf{Metric} \\ \midrule
Age heterogeneity  &
  Self-reported ages of team members &
  Normalized Coefficient of variation \\
Gender heterogeneity  &
  Participants self-reported their gender identity as “Male,” “Female,” or “Non-binary.” &
  Normalized Blau Index \\
Race heterogeneity  &
  Participants self-reported their race as “American Indian,” “African American or Black,” “Asian,” “White,” “Multiple races,” or “Other.” &
  Normalized Blau Index \\
International heterogeneity  &
  We asked participants, “Which country are you affiliated with?”, and they answered using a list of countries provided by the system. We binarized their answers and considered people identified with the US as “domestic,” and people identified with other countries as “international.” &
  Normalized Blau Index \\
Project skills heterogeneity  &
  Participants self-reported their expertise on six project skills relevant to the course using a 5-point Likert scale ranging from “Not at all skilled” to “Extremely skilled.” This metric was calculated by first calculating the coefficient of variation for each project skill, then summing the six normalized coefficients. &
  Normalized Coefficient of variation \\ \bottomrule
\end{tabular}%
}
\caption{Team composition’s attributes}
\label{tab:composition}
\end{table}

%% file: post_assembly_survey.tex
\begin{table}[!htb]
\small
\renewcommand{\arraystretch}{2}
\resizebox{\columnwidth}{!}{%
\begin{tabular}{>{\raggedright}p{0.15\textwidth}>{\raggedright}p{0.25\textwidth}>{\raggedright}p{0.35\textwidth}p{0.10\textwidth}p{0.1\textwidth}}
\toprule
\textbf{Dimension} &
  \textbf{Scale} &
  \textbf{Item/Question Example} &
  \textbf{Citation} &
  \textbf{$\alpha$} \\ \midrule
Usability &
  5-Point Likert Scale, 10 items &
  \textit{``I think that I would like to use this system frequently''} &
  \cite{Brooke1996-xx} &
  0.86 \\
Satisfaction with Team &
  5-Point Likert Scale, 4 items &
  \textit{``I am satisfied with the team I am in''} &
  AD &
  0.69 \\
Team self-efficacy &
  5-Point Likert Scale, 4 items &
  \textit{``My team has the right people to do a great job on this project''} &
  AD &
  0.83 \\
Motivation with the Team &
  5-Point Likert Scale, 3 items &
  \textit{``I am enthusiastic about working in this team''} &
  AD &
  0.92 \\
Connections efficiency &
  5-Point Likert Scale, 7 items &
  \textit{``In comparison to my experience forming teams, using the recommendation system to form a team… Connected me to a wider range of people''} &
  AD &
  0.85 \\
Prompting collaborators &
  5-Point Likert Scale, 7 items &
  \textit{``In comparison to my experience forming teams, using the recommendation system to form a team… Prompted me to think more carefully about who to work with''} &
  AD &
  0.92 \\
Perceived Differences - Surface Level &
  5-Point Likert Scale, 4 items (i.e., gender, age, race, ethnicity) &
  \textit{``How different are the members of your work in…?''} &
  \cite{Harrison2002-xd} &
  0.66 \\
Perceived Differences - Deep Level &
  5-Point Likert Scale, 5 items (i.e., personal values, project skills, backgrounds, personalities, education) &
  \textit{``How different are the members of your work in…?''} &
  \cite{Harrison2002-xd} &
  0.77 \\
Perceived recommendation quality &
  5-Point Likert Scale, 7 items &
  \textit{``The recommended teammates were well-chosen''} &
  \cite{Knijnenburg2012-ek} &
  0.88 \\
Perceived recommendation variety &
  5-Point Likert Scale, 5 items &
  \textit{``The recommendations contained a lot of variety''} &
  \cite{Knijnenburg2012-ek} &
  0.78 \\
Perceived trust &
  5-Point Likert Scale, 5 items &
  \textit{``I trusted the recommendations given by the recommender system''} &
  \cite{Knijnenburg2012-ek} &
  0.82 \\
Choice Satisfaction &
  5-Point Likert Scale, 7 items &
  \textit{``I am excited about my chosen teammates''} &
  \cite{Knijnenburg2012-ek} &
  0.78 \\
Privacy concerns &
  5-Point Likert Scale, 5 items &
  \textit{``I am afraid the recommendation system disclosed private information about me''} &
  \cite{Knijnenburg2012-ek} &
  0.83 \\
Distrust of algorithms &
  5-Point Likert Scale, 4 items &
  \textit{``I am less confident when I use algorithms''} &
  \cite{Knijnenburg2012-ek} &
  0.72 \\ \bottomrule
\end{tabular}%
}
\caption{Post-assembly survey items. Note: AD = author-developed items created for this study.}
\label{tab:post-assembly}
\end{table}

%% file: final_survey.tex
\begin{table}[!htb]
\centering
\small
\renewcommand{\arraystretch}{2}
\resizebox{\columnwidth}{!}{%
\begin{tabular}{>{\raggedright}p{0.15\textwidth}>{\raggedright}p{0.25\textwidth}>{\raggedright}p{0.35\textwidth}p{0.10\textwidth}p{0.1\textwidth}}
\toprule
\textbf{Dimension} &
  \textbf{Scale} &
  \textbf{Item/Question Example} &
  \textbf{Citation} &
  \textbf{$\alpha$} \\ \midrule
Communication &
  Sociometric question, participants marked the teammates they communicated. &
  \textit{“Who did you communicate with frequently?”} &
  AD &
  N/A \\
Psychological Safety &
  7-Point Likert Scale, 7 items &
  \textit{“Members of this team are able to bring up problems and tough issues”} &
  \cite{Edmondson1999-wk} &
  0.73 \\
Perceived Differences - Surface Level &
  5-Point Likert Scale, 4 items (i.e., gender, age, race, ethnicity) &
  \textit{“How different are the members of your work in…?”} &
  \cite{Harrison2002-xd} &
  0.63 \\
Perceived Differences - Deep Level &
  5-Point Likert Scale, 5 items (i.e., personal values, project skills, backgrounds, personalities, education) &
  \textit{“How different are the members of your work in…?”} &
  \cite{Harrison2002-xd} &
  0.80 \\
Team Processes (10 mini-scales) &
  5-Point Likert Scale, 30-item version. Main categories: Transition, Action, and Interpersonal Processes. &
  \textit{“To what extent did your team actively work to…”} &
  \cite{Marks2001-ha} &
  0.73 the lowest \\
Expertise Coordination &
  5-Point Likert Scale, 11 items &
  \textit{“The team has a good "map" of each other’s talents and skills”} &
  \cite{Faraj2000-de} &
  0.88 \\
Team Satisfaction &
  5-Point Likert Scale, 3 items &
  \textit{“Taken as a whole, I was satisfied with my team.”} &
  \cite{Peeters2006-sw} &
  0.94 \\
Team Trust &
  5-Point Likert Scale, 4 items &
  \textit{“Our team had a sharing relationship. We could freely share our ideas, feelings, and hopes.”} &
  \cite{McAllister1995-dh} &
  0.88 \\
Team Viability &
  5-Point Likert Scale, 4 items &
  \textit{“I really enjoyed being part of this team.”} &
  \cite{Bayazit2003-ry} &
  0.86 \\ \bottomrule
\end{tabular}%
}
\caption{Final survey items. Note: AD = author-developed items created for this study.}
\label{tab:final_survey}
\end{table}